\documentclass[12pt,preprint]{aastex}

\newcommand{\kms}{\rm km s$^{-1}\ $}

\newcommand{\heir}{\ion{He}{1}~$\lambda 10830$}
\newcommand{\heopt}{\ion{He}{1}~$\lambda 5876$}

\defcitealias{BEK}{BEK}
\defcitealias{HEG}{HEG}

\begin{document}

\slugcomment{Accepted by the Astrophsical Journal }
\title{Modeling T Tauri Winds from \heir\ Profiles}

\author{John Kwan \altaffilmark{1},
Suzan Edwards \altaffilmark{2},
William Fischer \altaffilmark{1} }

\altaffiltext{1} {Five College Astronomy, University of
 Massachusetts, Amherst, MA, 01003,
 kwan@astro.umass.edu, wfischer@astro.umass.edu}
\altaffiltext{2}{Five College Astronomy, Smith College,
 Northampton, MA 10163, sedwards@smith.edu}

\begin{abstract}
The high opacity of \heir\ makes it an exceptionally sensitive
probe of the inner wind geometry of accreting T Tauri stars. In
this line blueshifted absorption below the continuum results
from simple scattering of stellar photons, a situation which is
readily modeled without definite knowledge of the physical
conditions and recourse to multi-level radiative transfer. We
present theoretical line profiles for scattering in two possible
wind geometries, a disk wind and a wind emerging radially from
the star, and compare them to observed \heir\ profiles from a
survey of classical T Tauri stars. The comparison indicates that
subcontinuum blueshifted absorption is characteristic of disk
winds in $\sim 30\%$ of the stars and of stellar winds in $\sim 40\%$.
We further conclude that for many stars the emission profile of
helium likely arises in stellar winds, increasing the fraction
of accreting stars inferred to have accretion-powered
stellar winds to $\sim 60\%$.
Stars with the highest disk accretion rates are more likely to
have stellar wind than disk wind signatures and less likely to
have redshifted absorption from magnetospheric funnel flows.
This suggests the possibility that when accretion rates are
high, disks can extend closer to the star, magnetospheric
accretion zones can be reduced in size and conditions arise that
favor radially outflowing stellar winds.

\end{abstract}


\section{INTRODUCTION}

Young stars with accretion disks possess
extended collimated mass outflows on spatial
 scales as small as a few AU
\citep{hep04,doug00} to as large as a few parsecs \citep{bally}.
These ubiquitous outflows are almost certainly
accretion-powered since there is a robust correlation between
mass accretion and outflow rates over many orders of magnitude
\citep{HEG,rich00}. The flows are thought to be launched via
magnetohydrodynamic processes, where a large scale open magnetic
field is anchored to a rotating object, but whether that object
is the disk or star or both is still uncertain \citep{jf06}.
Disk winds may either emanate over a wide range of disk
radii \citep{kp00} or be restricted to a narrow (X) region near
the disk truncation radius \citep{shu94}, while stellar winds
follow open field lines emerging radially from the star
\citep{hart82,fc05,rb05,love05}. Disk winds will aid in angular
momentum transport in the disk and carry angular momentum
away from the system \citep{pud06}, thereby affecting the
evolution of the disk and any planets that may form there.
Stellar winds in accreting systems are often regarded as of
secondary importance, but recent work by Matt and Pudritz (2005)
suggests otherwise. After raising serious questions as to
whether a magnetic interaction between the disk and the star can
provide the necessary spin-down for accreting stars, they posit
that it is accretion-powered stellar winds that counteract the
spin-up torque acquired by a star during disk accretion
\citep{hart89} and are thus the primary agent for
regulating stellar angular momentum in forming stars.

Observational constraints on the wind launch region are thus crucial
for clarifying outflow origins and providing insight into the
evolution of angular momentum in accretion disks and accreting
stars, yet direct empirical tracers have been surprisingly
elusive. No evidence for winds is seen in molecular lines of
$CO$ and $H_2O$ that arise in the inner accretion disk at radii
from 0.1 to 2 AU. Rather, these lines show disk profiles that
include turbulent broadening and warm disk chromospheres
\citep{naj03,carr04}, and their symmetry belies no trace of
outflowing gas from the upper disk layers. In contrast atomic
lines that form within 0.1 AU of the star, in particular
H$\alpha$, Na D, Ca II H\&K, and MgII h\&k \citep{app89,naj00,ard02},
show clear evidence for inner winds at velocities of several
hundred \kms\ in many classical T Tauri stars (CTTS) in the form
of blueshifted absorption features superposed on broad emission.
However, their utility as a wind diagnostic is limited since the
dominant contributor to the profile is the strong, broad, and centrally
peaked emission that is generally interpreted as arising in
magnetospheric infall zones \citep{mch01}.

Recently, a new and quite robust diagnostic of the inner wind in
accreting stars has been recognized. In contrast to the hydrogen
lines, \heir\ profiles in CTTS and Class I sources show P
Cygni-like profiles with deep blueshifted absorption below the
continuum \citep{e03,dup05}, reminiscent of those in stellar
wind models developed to account for CTTS Balmer lines over a
decade ago \citep{hart90}. It was in fact the inability of wind
models to account for the structure of CTTS Balmer lines that
led to the re-interpretation of these lines as arising in
magnetospheric infall zones, which met with greater success in
describing hydrogen emission profiles \citep{hhc94,e94}.
The extraordinary potential for \heir\ to diagnose
inner winds is revealed in a recent survey of classical T Tauri
stars where profiles show subcontinuum
blueshifted absorption in $\sim$70\% of the stars,
in striking contrast to H$\alpha$ where only
 $\sim$10\% of CTTS have blue absorption penetrating the
continuum \citep [hereafter EFHK]{e06}. The breadth and
penetration depth of this subcontinuum blue absorption make
\heir\ an unprecedented probe of the wind geometry and
acceleration region, providing for the first time the potential for good
constraints on the launching region of the high velocity inner
wind.

In this paper we look more deeply into the conclusion put forward
in EFHK, that the diversity in the morphology of the
subcontinuum blueshifted absorption at \heir\ requires a range
of inner wind geometries in accreting stars. Many CTTS have P
Cygni-like profiles suggesting stellar winds while others have
narrow blue absorptions suggesting disk winds. EFHK also find
that both kinds of winds are likely accretion-powered as no
evidence for winds is seen at $\lambda 10830$ in non-accreting
weak T Tauri stars (WTTS)
while among accreting CTTS the combined strength of helium
emission plus absorption is well correlated with the level of
simultaneously measured $1\mu m$ veiling, $r_Y$, defined as the
ratio of continuum excess to the photospheric flux. The
existence of helium emitting winds in CTTS was first suggested
in a study of optical He I $\lambda 5876$ profiles
\citep[hereafter BEK] {bek}, which in contrast to \heir\ show
several emission components but no subcontinuum blueshifted
absorption. Although these two helium transitions are intimately
connected -- the $\lambda 5876$ transition lies immediately
before $\lambda 10830$ in a recombination cascade sequence --
the difference in their line profiles arises from the
metastability of the lower level of He I $\lambda 10830$.
While $\lambda 10830$ is optically thick, the absence of blue
absorption at $\lambda 5876$ indicates it traces the wind with
optically thin emission. The contrast in optical depth for this
pair of closely coupled lines thus offers a particularly potent
means of diagnosing T Tauri winds.

Here, we examine the origin of \heir\ profiles in accreting stars
by calculating theoretical profiles produced individually by
both disk and stellar winds and comparing them to those from the
EFHK survey. The frequent presence of absorption features in
this line is due to its high opacity, owing to the nature of the
$2p^3P^o - 2s^3S$ transition. The metastable lower level
($2s^3S$), $21~eV$ above the singlet ground state, will be
significantly populated relative to other excited levels owing
to its weak de-excitation rate via collisions to singlet states.
And as its upper level ($2p^3P^0$) has only $\lambda 10830$ as a
permitted transition, absorption of a stellar continuum photon
by this state will lead to re-emission of a $\lambda 10830$
photon, unless the electron density is sufficiently high that
collisional de-excitation ensues. We do not consider the latter
regime here, and we model the $\lambda 10830$ line from the
resonant-scattering of continuum photons. The resulting line
profile depends on the line opacity but is insensitive to its
value once it exceeds unity, so an examination of how the
profile morphology depends on the geometrical structure of the
gas flow can be obtained quite readily by considering the case
in which the gas flow is opaque to the stellar continuum at each
velocity. These theoretical profiles for disk and stellar winds
are presented for cases of pure scattering and scattering plus
in-situ emission in \S 2. Their characteristics are summarized
and compared with observed profiles, including both \heir\ and
\heopt, in \S 3. The discussion in \S 4 addresses the relation
between winds and accretion flows and the likely simultaneous
presence of disk and stellar winds. A conclusion of our findings
is given in \S 5.

\section{THEORETICAL LINE PROFILES}

We present theoretical line profiles for two possible wind
geometries, a disk wind and a wind emerging radially from the
star. Both models predict profile morphologies (1) for pure
scattering of stellar photons by the wind (absorption and
subsequent reradiation) and (2) for scattering plus an
additional source of in-situ emission in the wind. A full
Monte Carlo approach is followed only for the stellar wind,
since, as explained in the following sections, the effects of
secondary scattering on the emergent profile will be minor for
the disk wind but significant for the stellar wind. Geometrical
effects explored for stellar winds are disk shadowing,
restricting the wind to emerge from stellar polar regions, and
restricting the origin of incoming photons to an axisymmetric
ring on the stellar surface, as might be expected for radiation
from a hot magnetospheric accretion shock. The resultant profile
morphologies are compared with observed \heir\ profiles for
classical T Tauri stars in Section 3.

\subsection{Disk Wind Line Profiles}

An azimuthally-symmetric disk wind is assumed to flow away from
the disk along streamlines tilted at an angle $\theta_w$ from
the normal to the disk plane. This wind opening angle is always
$\theta_w > 30^o$, as required for magnetocentrifugal launching
\citep{bp}. The wind is launched between an inner radius
$\rho_i$, at which the disk is truncated, and an outer radius
$\rho_f$, where $\rho$ is the distance along the disk plane from
the star measured in units of $R_*$, the stellar radius. It is
reasonable to expect the terminal poloidal velocity of the gas
launched from $\rho$ to scale with the escape velocity, which is
$\propto \rho^{-1/2}$, and the distance to reach this terminal
value to scale with $\rho$. We thus assume a simple poloidal
velocity law that encompasses these relations with
$v_p=540\rho^{-1/2} (1-\rho/r)^{1/2} $ \kms, where $r$ is the
radial distance from the star, also in units of $R_*$, and the
scaling factor establishes the maximum terminal velocity (for
$\rho_i$ = 2) as 382 \kms, in line with observed values. The
exact details of the gas acceleration are not crucial, so long
as the poloidal velocity increases monotonically from zero to
the terminal value, because we also assume the disk wind to be
opaque in the scattering transition (representing \heir) at each
position. This assumption on the opacity ensures, for an optimal
comparison with the stellar wind models, the maximum breadth and
depth of the continuum absorption that can be produced by a disk
wind scattering stellar photons. If the line becomes optically
thin at some distance away from the disk plane, which is
possible and likely, the resultant absorption from a disk wind
will be weaker, particularly at the high
velocity end, than the model one.

The angular velocity of the disk where each streamline is anchored is
specified by $\Omega=(300~km~s^{-1}/R_*)\rho^{-3/2}$. We examine the
effects of rotation in the disk wind for two extreme cases. In one
case each individual streamline rotates at a rate that conserves
angular momentum. In the other case each streamline rotates
rigidly. These two extremes should encompass the realistic case
of rigid rotation close to the disk and rotation with
conservation of angular momentum farther away.

With the assumption of a completely opaque scattering
transition the absorption component
of the profile simply reflects the contour area projected by the disk
wind against the star at each line-of-sight velocity.
 An absorbed stellar photon is assumed to be re-emitted
isotropically and right away it either escapes to infinity or
hits the star or disk. In contrast to the stellar wind
simulations in the following section we have not followed the
detailed photon path of multiple scatterings since in this case
the scattered photons will not appreciably alter the morphology
of the narrow absorption component. This is because in a disk
wind more photons are scattered out of a viewing angle
intersecting the wind than into the same viewing angle, and
those that are scattered into the same viewing angle are very
much broadened by the large bulk velocity (rotational and
poloidal motions) of the flow, so they are widely distributed in
velocity.

\begin{figure}
\epsscale{0.8}
\plotone{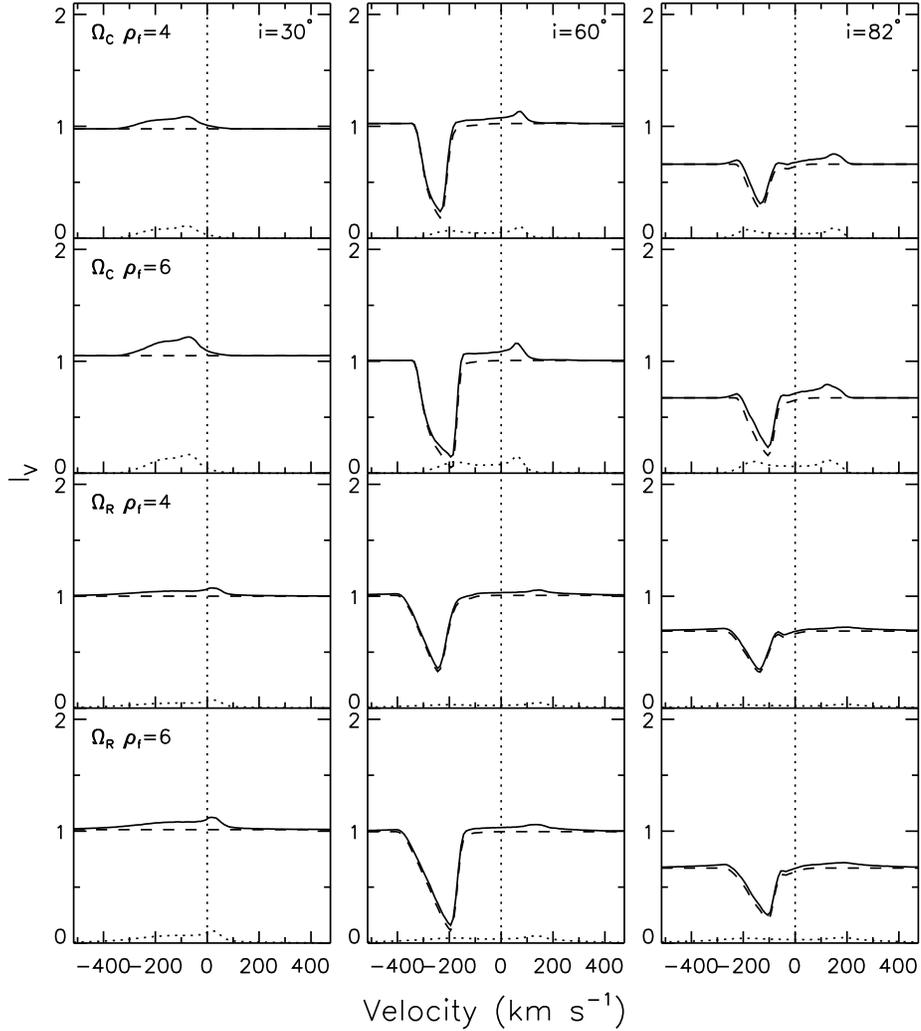}
\figcaption{DISK WIND: Theoretical line profiles produced from
scattering of the stellar continuum by a disk wind emanating at
$45^o$ from the disk normal. Within each window the dashed line
shows the profile of the absorbed stellar continuum, the dotted
line the profile of the scattered photons, and the solid line
the summed emergent profile. The three columns denote, from left
to right, viewing angles $i=30^o$, $60^o$, and $82^o$. The top
two rows illustrate the case in which the disk wind rotates with
conservation of angular momentum, $\Omega_C$, and the bottom two
rows the case of rigid rotation, $\Omega_R$. In each rotation
case the upper row has the disk wind originating from between 2
to 4 $R_*$, and the lower row from 2 to 6 $R_*$.
\label{f.disk}}
\end{figure}

\begin{figure}
\epsscale{0.8}
\plotone{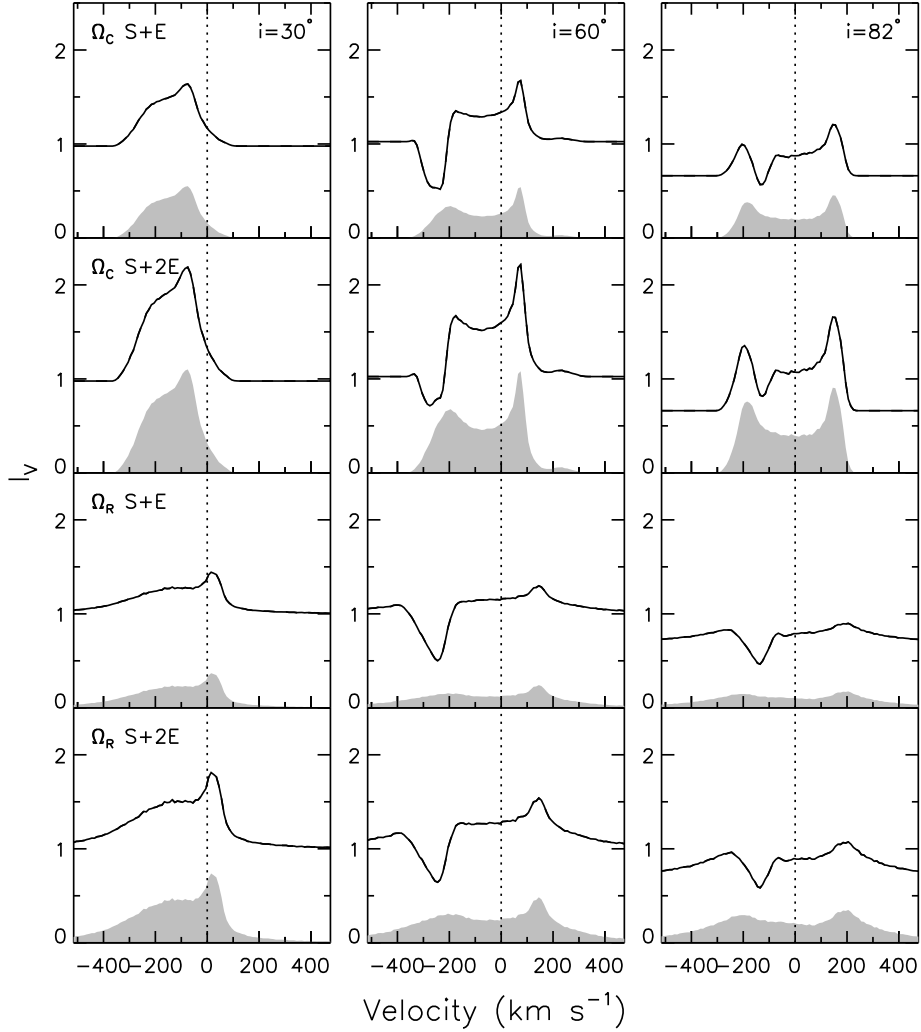}
\figcaption{DISK WIND: Theoretical line profiles from a disk wind
emanating at $\theta_w =45^o$ with in-situ emission added to
pure scattering. Shaded areas show the in-situ contribution,
scaled at 5 times the scattered emission ($5S$) in rows 1,3  and
10 times the scattered emission ($10S$) in rows 2,4 and the solid
line is the emergent profile. As in Figure~\ref{f.disk}, columns
reflect viewing angles from $i=30^o-82^o$, top 2 rows are rotation
with conservation of angular momentum $\Omega_C$, and bottom two
are rigid rotation, $\Omega_R$. All cases have $\rho_i$=2,
$\rho_f$=4.
\label{f.disk2}}
\end{figure}

Predicted line profiles from disk winds are shown in
Figures~\ref{f.disk} and \ref{f.disk2}, with model parameters
listed in Table 1. The first figure is for pure scattering and
the second figure includes additional contributions from in-situ
emission (see below). In both figures the wind opening angle is
$\theta_w =45^o$ and $\rho_i=2 $. In Figure~\ref{f.disk} cases
are shown for both $\rho_f=4$ and $6$, but only for $\rho_f=4$
in Figure~\ref{f.disk2}. Columns show 3 viewing angles:
$i=30^o$, $60^o$, and $82^o$. The top two rows show the case of
conservation of angular momentum ($\Omega_C$), and the bottom
two the case of rigid rotation ($\Omega_R$). The solid line is
the resultant profile, which is a sum of the absorption of the
continuum (dashed line), the emission from scattered photons
(dotted line), and in-situ emission (shaded area), if present.
Slight deviations from unity in the continuum level
 at $i=30^o$ and $60^o$ are statistical
fluctuations caused by the finite number of photons used in
the simulation.  The much lower continuum level at
$i=82^o$ is due to disk occultation.

For the pure scattering case in Figure~\ref{f.disk}, at
viewing angles less than the opening angle of the wind,
$i<45^o$, the resultant profiles are
purely in emission and almost entirely blueshifted, owing to
disk occultation of the receding wind. They are also quite broad
($>$ 300 \kms), more so for the case of rigid rotation because
of the larger rotational broadening. For viewing angles
exceeding the opening angle of the wind, $i>45^o$, the line of
sight to the star crosses the wind streamlines and the profile
is prominently in absorption. The emission contributed from the
scattered photons is weak compared to the absorption because
there is no scattering into $i>\theta_w$ from continuum absorption
at $i<\theta_w$ and because of disk occultation.

{\it The distinctive aspect of absorption features from a disk
wind is that they are narrow in comparison to the full range of
velocities in the accelerating wind, never extending over the
whole possible range from rest to terminal velocity despite the
assumption of high opacity}.  This is because at a particular viewing
angle most lines of sight toward the star will intercept only a
small portion of the full range of wind velocities; those toward
the stellar polar region intersect the wind near its terminal
speed while those toward the stellar equatorial region intersect
the wind at low velocities. Thus in the summed emergent profile
only a small range of projected velocities will be encountered
by sufficient lines of sight to produce a significant
absorption. Similar reasoning also accounts for a shift in the
centroid of the absorption feature toward a smaller blueshift as
$i$ increases and the earlier part of the acceleration region is
preferentially intercepted.  The absorption does not bottom out
because at no observed velocity is the stellar continuum
completely covered.

In addition to their dependence on viewing angle, the width and
centroid of the absorption component will also be affected by
the range of radii in the disk over which the wind is launched.
For our inner launch radius of $\rho_i =2$, we explore outer
launch radii from $\rho_f=4$ to $\rho_f=6$. Although these
maximum launch radii are smaller than the full range expected
for an extended disk wind, this choice is appropriate for the
region of helium excitation, which will be restricted to the
inner disk, which is also where the high speed component of jets
will arise. It can be seen by comparing the first row
($\rho_f=4$) with the second row ($\rho_f=6$) for each rotation
case in Figure~\ref{f.disk} that the blue absorption extends
toward lower velocities when $\rho_f$ is increased and more
streamlines of comparatively lower velocities are seen through.
The effect of changing $\rho_i$ can be seen from difference
profiles between the first and second rows, where a narrower
width and less blue centroid would arise from a wind launched
between $\rho_i=4$ and $\rho_f=6$. Additionally, the line width
will be affected by rotation, as shown for the two extreme cases
we have examined, where the absorption has a slightly larger
width when the disk wind rotates rigidly ($\Omega_R$).

Changing $\theta_w$ will not appreciably alter the profile
characteristics. A wider opening angle at the same $i$ shifts
the centroid of the absorption  blueward and makes
it slightly broader. This results because the streamline is then
more closely aligned with the viewing angle, and because the gas
speed is higher at the intersection point (larger value of
$r/\rho$ in the expression for $v_p$). Wider opening angles also
reduce the probability of observing an absorption profile, which
requires $i>\theta_w$. Narrower opening angles will reverse
these effects, but must always be $\theta_w > 30^o$.

The effect of including a contribution from in-situ emission in
the disk wind is shown in Figure~\ref{f.disk2}. We simply assume
that the in-situ emission line profile has a kinematic structure
identical to the profile of scattered photons, since its main
characteristics come from the large bulk velocity at each point
of the wind from the combined rotational and poloidal motions,
and the line profile will depend only weakly on the excitation
details. To illustrate its effect on the profile, we adopt a
value for the in-situ contribution that is comparable to the
magnitude of the absorption. In this case the in-situ emission
is 5 times the scattered emission ($E=5S$). We also include a
case where it is twice as strong as the absorption ($E=10S$). As
in Figure~\ref{f.disk}, the first 2 rows of Figure~\ref{f.disk2}
are for the case of conservation of angular momentum and the
last two for rigid rotation, where now the first row in each
case is the smaller, and the second the larger, contribution
from in-situ emission. It is seen that the effect of this
additional source of emission does not significantly alter the
profile morphology. Although the broad, flat emission is more
prominent than before, the sharp blue dip will still be visible
because the large bulk motions distribute the emission broadly
over all velocities. At viewing angles inside the wind opening
angle, where no absorption is formed, the broad blueshifted
emission is now quite prominent. As the viewing angle increases,
the blue absorption becomes apparent, still narrow compared with
the extent of the emission and decreasing in velocity as the
viewing angle increases. At the same time, the emission shifts
to a more centered but very broad profile, clearly showing the
distinctive double-horned structure characteristic of rotation.

The above profile characteristics of disk winds are inherent to
the location of the wind relative to the star and the geometry
of the streamline, with a strong aspect dependence depending on
whether the viewing angle is interior or exterior to the wind
opening angle. Our assumption of a completely opaque scattering
transition ensures that the model profiles show the
maximum possible blueshifted absorption from a disk wind. In the
next section we examine the case of a stellar wind in which the
geometry of the acceleration process, being radially away from
the center of the star, enables each line of sight toward the
center of the star to see through the full velocity range from
$0$ to the terminal value.

\subsection{Stellar Wind Line Profiles}
 
We first examine the simple case of a spherical, radial stellar wind
originating from an inner radius $r_i$ and having a velocity law
given by $v=v_t (1-r_i/r)^{1/2}$, with $v_t$ being the terminal
velocity. This velocity law is analogous to the one used
for the disk wind, except that there is only one origination
radius. Our models also allow for the presence of a geometrically
flat disk, with a truncation radius specified by $\rho_T$.

The stellar wind calculation differs from that of the disk wind
in several ways. First, we relax the assumption of a completely opaque
transition, and second, we employ a Monte Carlo,
rather than single scattering, simulation of the radiative transfer.
The first change is made to accomodate the observed profiles in the
EFHK survey with a P Cygni-like character that show blue
absorptions tapering toward the continuum at the highest
velocities, indicating that the line is becoming optically thin.
We assume a simple radial dependence for the density in the
lower state of the scattering transition representing \heir,
i.e., $n_{2^3S}\propto (r_i/r)^\alpha v_l/(v+v_l)$, where
$\alpha$ is a parameter and $v_l$, taken to be $30~km~s^{-1}$,
avoids divergence at $v=0$. The line opacity as a function of velocity
is determined by a suitable choice of $\alpha$, so that
the line opacity ranges from being much greater than $1$ at
low velocities, thereby producing a trough-like absorption, to
less than $1$ at high velocities, thereby producing a tapering
towards the continuum. The outflow is also given a turbulence
velocity width given by $\Delta v_{tb}=\Delta v_i(r_i/r)^\beta$,
with $\Delta v_i$ and $\beta$ being parameters. Two cases are
considered, a low turbulence one ($T_L: \Delta v_i=10~km~s^{-1}$
and $\beta=0$) and a high turbulence one ($T_H:\Delta
v_i=100~km~s^{-1}$ and $\beta=2$). The extreme value for the high
turbulence is chosen because one observed CTTS profile (DR Tau)
shows a broad blue
absorption extending redward of the stellar velocity by up to
$100~km~s^{-1}$. For the low and high turbulence cases $\alpha$
is taken to be 6 and 5 respectively.

The adoption of a full Monte Carlo approach is necessary here because,
in contrast to the disk wind case, the absorption of the stellar
continuum by a stellar wind will be substantially filled in by
scattered photons. This requires that we obtain the profiles with a
Monte Carlo simulation so the full path can be traced for each photon
until it either escapes to infinity or hits the star or the
disk.

As in the disk wind, we examine the effects of allowing the wind
gas to produce in-situ emission in addition to scattering of the
stellar continuum, representing the possibility of the
additional production of \heir\ photons via recombination and
cascade and/or collisional excitation. In this case however the
in-situ emission profile is derived from a Monte Carlo
simulation, parameterized with a rate $\propto
n_{2^3S}(r_i/r)^\gamma$. The effect of changing $\gamma$ from
$0$ to $2$, corresponding to successively smaller contributions
from the outer, higher-speed gas has a small effect on the
emitted profiles, and the ones shown are for $\gamma =1$. As for
the disk wind simulation, we choose the magnitude of the in-situ
emission ($E$) to be comparable to the absorption strength of
the stellar continuum, so that the summed line profile ($S+E$)
will illustrate an intermediate case between pure scattering and
dominant in-situ emission.

\begin{figure}
\plotone{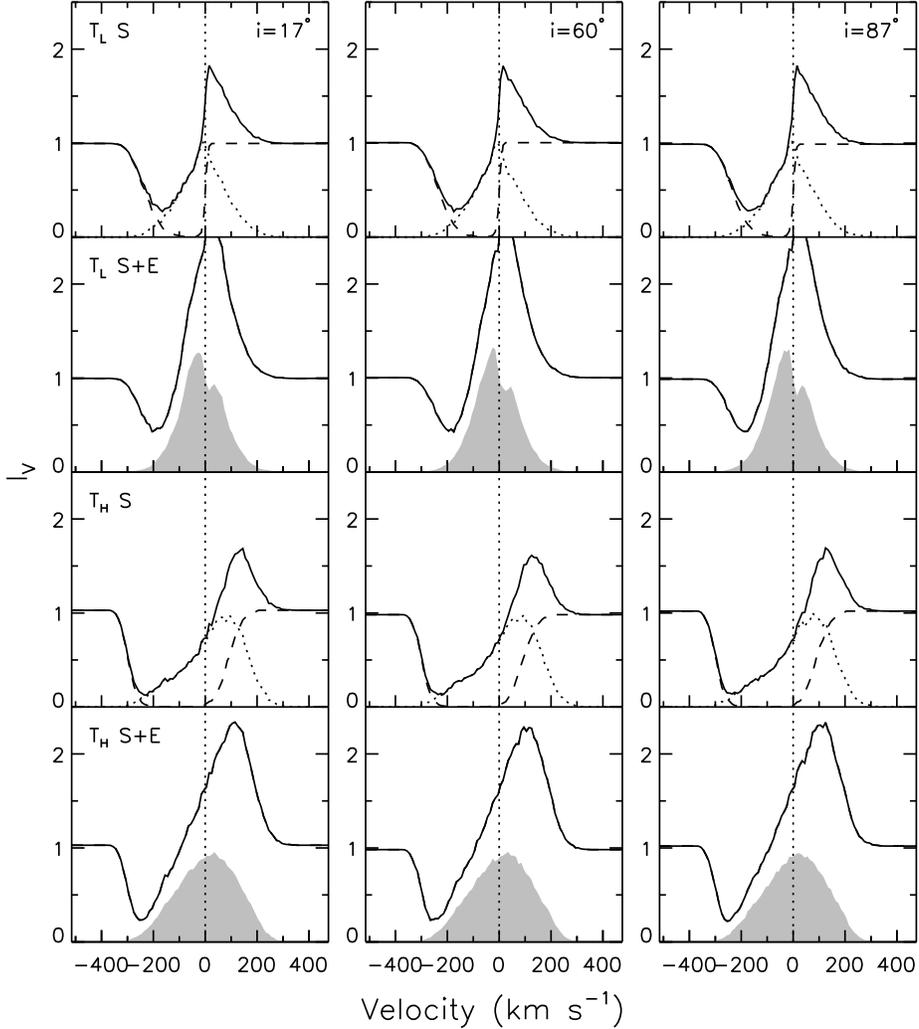}
\figcaption{ STELLAR WIND: Theoretical profiles obtained by
Monte Carlo simulation of line formation in a radially expanding
spherical wind originating at $r_i$ = 1.5 $R_*$ in
the absence of disk occultation ($\rho_T>5r_i$ ). The top two
rows represent the low turbulence case, $T_L$, and the lower two
the high turbulence case, $T_H$. For each turbulence case, the
first row is for pure scattering of the stellar continuum ($S$)and the
second includes additional in-situ emission ($S+E$). The shaded regions
show the in-situ contribution, scaled to be comparable in
strength to the scattering absorption.
Although there is no viewing angle dependence in the
absence of a disk, the figure is formatted for ready comparison with
the remaining figures, and vertical columns
denote 3 viewing angles, $i=17^o$, $60^o$, and $87^o$ measured from the
disk normal.
\label{f.stellarw_nodisk}}
\end{figure}

\begin{figure}
\plotone{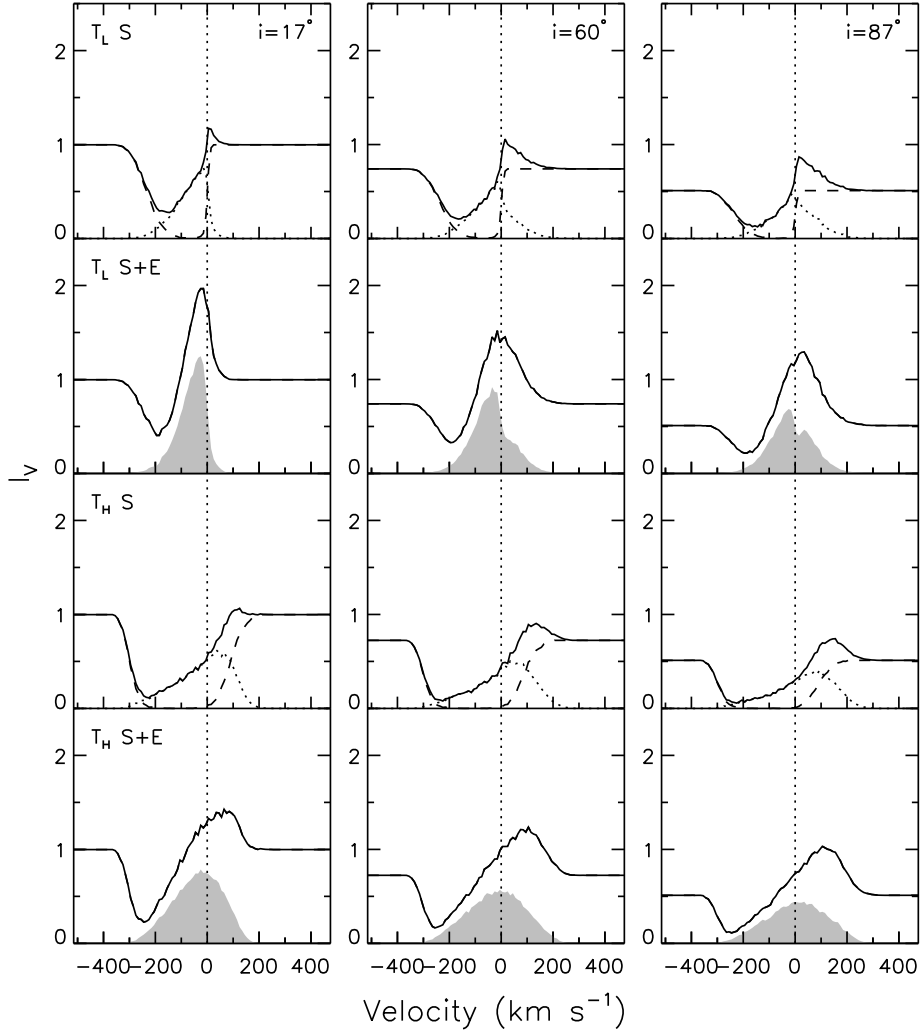}
\figcaption{STELLAR WIND: Same spherical wind as in
Figure~\ref{f.stellarw_nodisk}, but now including the effects of
disk shadowing (important when $\rho_T \le r_i$). Here the disk
reaches to the stellar surface, $\rho_T = 1 R_*$ and $r_i$ = 1.5
$R_*$. For each turbulence case ($T_L$, $T_H$) profiles without
and with in-situ emission contributions ($S$, $S+E$) are shown.
Vertical columns denote 3 viewing angles, $i=17^o$, $60^o$, and
$87^o$, measured from the disk normal.
\label{f.stellarw_disk}} \end{figure}

\begin{figure}
\plotone{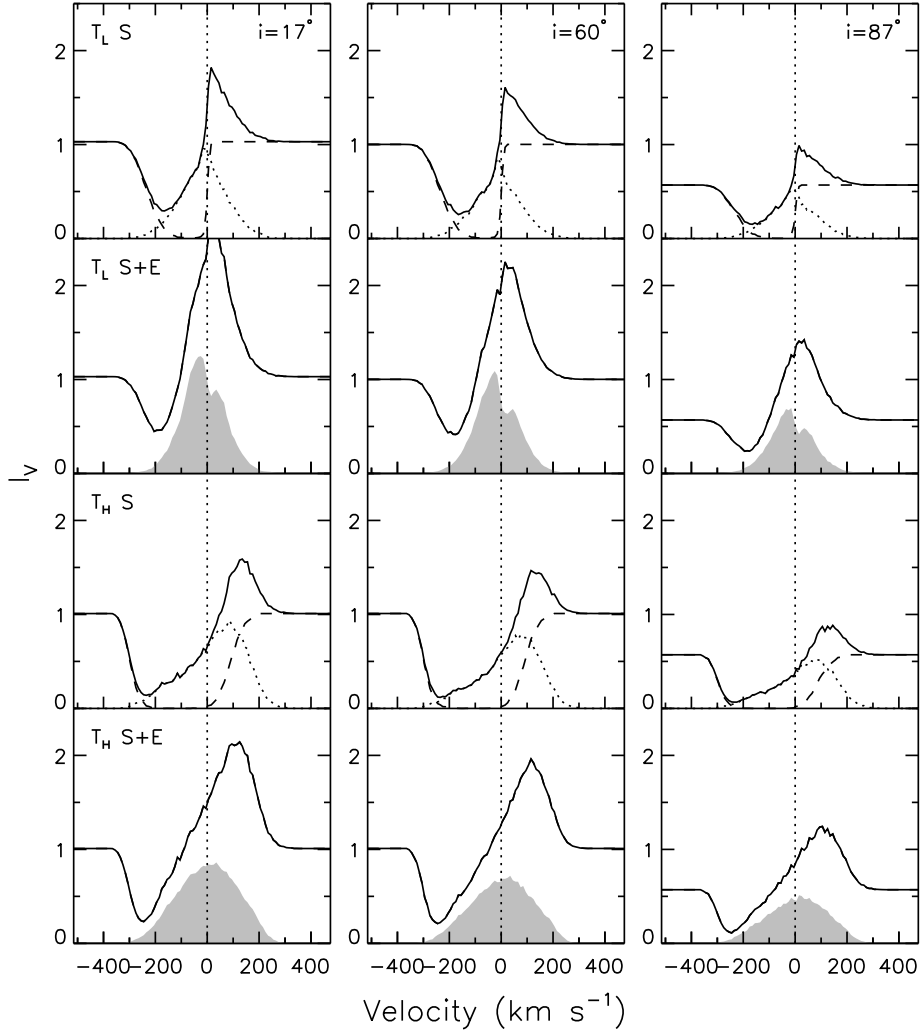}
\figcaption{STELLAR WIND: Same as Figure~\ref{f.stellarw_nodisk},
but in the presence of a disk truncated at $\rho_T =2.25 R_*$.
For each turbulence case ($T_L$, $T_H$) profiles without and with in-situ
emission contributions are shown ($S$, $S+E$). Vertical columns denote 3
viewing angles, $i=17^o$, $60^o$, and $87^o$, measured from the
disk normal. \label{f.stellarw_trdisk}}
\end{figure}

Figures~\ref{f.stellarw_nodisk},\ref{f.stellarw_disk},\ref{f.stellarw_trdisk}
illustrate profiles for a spherical stellar wind originating at $r_i=1.5R_*$,
with 3 different disk truncation radii and 4 different sets of wind assumptions,
as summarized in Table 2. This set of models shows clearly
the effect of disk shadowing, which depends on the
relative sizes of the wind origination and disk truncation radii.
There will be no shadowing effect when the disk truncation
radius is large compared to the wind origination radius,
$\rho_T>5r_i$, while they will be significant
when $\rho_T \le r_i$. Thus it is the ratio $\rho_T/r_i$ that
determines the disk shadowing effects. Profiles for
a particular ratio would be the same were it not for the additional
effects of stellar occultation, which will be more important when $r_i$
is closer to $R_*$.

Figure~\ref{f.stellarw_nodisk} is for a spherical wind with no
disk. Although accretion disks are present in all the classical
T Tauri stars whose helium profiles we are modeling, this set of
wind profiles would apply to those disks where disk shadowing is
not important, i.e. where the inner truncation radius is large
compared to the starting point for the wind, $\rho_T>5r_i$. The
case of extreme disk shadowing is shown in
Figure~\ref{f.stellarw_disk} for a disk reaching to the stellar
surface ($\rho_T=1$). However, similar profiles would result as
long as the wind origination radius much exceeds the disk
truncation radius, $\rho_T\le 0.5~r_i$.
Figure~\ref{f.stellarw_trdisk} illustrates an intermediate case
where the disk is truncated at a distance that is only 1.5 times
the starting point for the wind, $\rho_T=1.5r_i$. In each figure
the 3 vertical columns represent views from more pole-on to more
edge-on configurations, $i=17^o$, $60^o$, and $87^o$. In the
diskless case of Figure~\ref{f.stellarw_nodisk} we present
profiles for the same 3 viewing angles to ensure similar
proportions among all the figures and facilitate comparison of
profile morphology with the cases including disk shadowing.
As expected in the absence of a disk there
is no dependence on $i$, although small effects from the finite
statistics in the Monte Carlo simulation can be seen on close
inspection.

The four horizontal rows in Figures~\ref{f.stellarw_nodisk},
\ref{f.stellarw_disk}, \ref{f.stellarw_trdisk} correspond to 4
different assumptions about the stellar wind. In all 4 cases the
wind originates at the same distance from the star, $r_i=1.5$,
and has the same terminal velocity, $v_t=375$ \kms, chosen to be comparable
to the maximum terminal velocity in the disk wind. Two cases
(top two horizontal rows) represent models with low turbulence
($T_L$) and the other two cases (bottom two horizontal rows)
models with high turbulence ($T_H$) plus a somewhat higher
optical thickness of the gas ($\alpha=5$ rather than $6$). For
each turbulence case, the profiles in the upper row are formed
simply from scattering of the 1 $\micron$ stellar continuum
photons, where the dashed line shows the absorption of this
continuum, the dotted one shows the profile of the photons
scattered into the line of sight, and the dark solid line is the
summed emergent profile. The lower row of each turbulence case
includes the effects of additional in-situ emission in the wind,
where the shaded region shows the contribution from in-situ
emission alone and the solid line is the emergent profile
resulting from summing this with the pure
scattering profile from the previous row.

The occultation effect of the disk on the line profiles is
manifest in three ways. One, theoretically significant but
difficult to establish observationally, is that the stellar
continuum is dependent on viewing angle. In the $\rho_T=1$ disk
the continuum level drops steadily from $1$ to $0.5$ as $i$
increases from pole-on ($0^o$) to edge-on ($90^o$), and in the
$\rho_T=1.5r_i$ disk it remains at 1 for $i<64^o$ and then drops
toward $0.5$ at $i=90^o$.

The other two effects directly impact the profile morphology,
altering the {\it net} equivalent width and the centroid of the
emission component. In the diskless case
(Figure~\ref{f.stellarw_nodisk}) the intrinsic wind profile from
pure scattering has a P-Cygni character, with prominent blue
absorption and red emission components. Its net equivalent width
is somewhat negative (i.e. absorption exceeds emission )
rather than zero because of stellar
occultation, and this effect is more pronounced in the case of
high turbulence (row 3) owing to the preferential back
scattering of the stellar photons. The presence of a disk has a
minimal occultation effect on the line profile for edge-on views
but this becomes significant for more pole-on views. Although
all 3 disk cases have similar pure scattering profiles after
normalization of the continuum level at $i = 87^o$, as $i$
decreases the net equivalent width for the $\rho_T=1$ disk
becomes more pronouncedly negative, and by $i=17^o$ there is
hardly any red emission, and the observed profile is entirely in
absorption. In the $\rho_T=1.5r_i$ disk the line profile at
$i=60^o$ is intermediate between those of the other two disks,
while at $i=17^o$ it becomes like that of the diskless case.
This is because, given the acceleration law, the gas reaches
$0.58 v_t$ at $r=1.5r_i$, so a large part of the photon
absorption occurs within this radius, and the hole in the disk
allows most of the photons scattered by the receding stellar
wind to come through at small inclination angles. {\it Hence a
large negative net equivalent width requires the disk truncation
radius $\rho_T$ be comparable to or smaller than the wind origination radius
$r_i$}.

The second observable effect of disk occultation is altering the
centroid of any in-situ emission. This is most evident in the low
turbulence case (2nd row from top). In the $\rho_T=1$ disk
(Figure~\ref{f.stellarw_disk}) at $i=87^o$ the in-situ emission
profile (shaded area) is nearly symmetric about $v=0$,
the stellar velocity. (The slight depression in the in-situ
emission immediately to the red of $v=0$ and the slightly
weaker red side is caused by stellar
occultation.) But, as $i$ decreases and more of the receding
stellar wind is occulted the emission profile becomes
progressively more blue-shifted. This characteristic for a line
purely in emission is often indicative of the
presence of outflowing gas in the presence of an obscuring disk.
When both scattering and in-situ emission contribute roughly
comparably, the emergent profile also reveals a blueshifted
emission centroid.  Thus the {\it emergent profiles
with in-situ emission show that, when disk shadowing is not
important, the emission component of the observed profile has a red
centroid, but the emission progresses from red to blue
centroids as disk shadowing becomes more effective.}

We included calculations with extreme turbulence because
the DR Tau profile shows a broad blue absorption
extending redward of the stellar velocity by up to
$100$ \kms. The high turbulence case we considered (row 3)
does successfully reproduce this effect. It also leads to a
preferential backscattering of the stellar photons, thereby
producing a redshifted centroid in the profile of the scattered
photons, which is clearly seen in the diskless case (dotted line
in Figure~\ref{f.stellarw_nodisk}). In the $\rho_T=1$ disk
(Figure~\ref{f.stellarw_disk}) it is most obvious at $i=87^o$,
but is still noticeable at $i=17^o$, despite disk occultation of
some of the red photons, in comparison with the low turbulence
case (row 1). The combination of these two effects of
turbulence, namely extending the blue absorption redward of the
stellar velocity and creating a redshifted scattering profile,
is a more distinct and well displaced red emission centroid in
the emergent profile for pure scattering. The backscattering is
much more important for the stellar photons, which propagate
into the wind along $\hat{r}$, than for the photons generated in
the wind which are emitted with an initial isotropic
distribution in direction (compare the dotted line in row 3 with
the shaded area in row 4). Instead, the effect of turbulence on
the in-situ emission components is to broaden them, while disk
occultation makes them more blue-shifted for more face-on
orientations. The combined scattering plus emission profile then
shows an emission centroid whose position is dependent on
viewing angle. (Here the in-situ emission profiles are shown
with $\gamma=1$; for $\gamma=0$ and $2$ the corresponding
profile will be similar, but slightly broader and narrower
respectively.)

\begin{figure}
\plotone{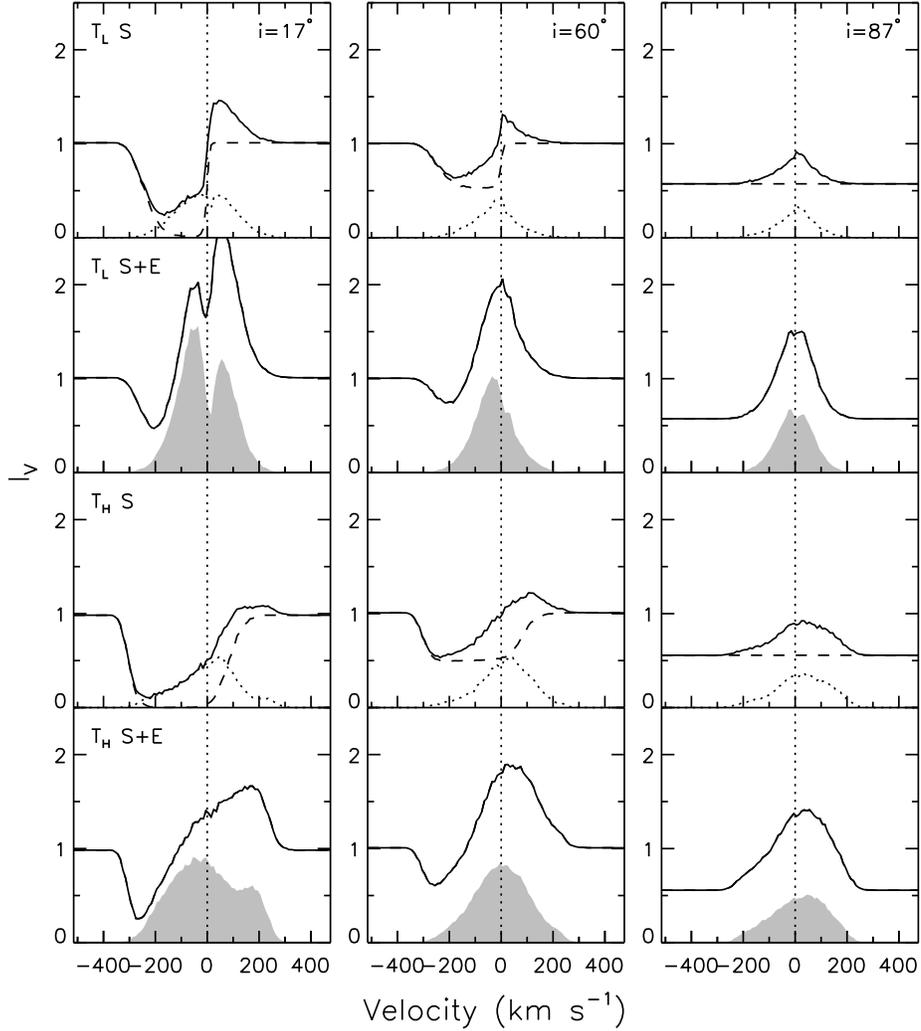}
\figcaption{POLAR STELLAR WIND: Profiles for a polar stellar wind
emerging within $60^o$ of the pole, for the same two cases of
turbulence ($T_L$, $T_H$) with and without in-situ emission
($S$, $S+E$) and including the effects of disk shadowing ($\rho_T \le r_i$),
where $r_i = \rho_T= 3.0 R_*$. As before, vertical columns denote 3 viewing
angles, $i=17^o$, $60^o$, and $87^o$, measured from the disk
normal.
\label{f.stellarw_pole}}
\end{figure}

\begin{figure}
\plotone{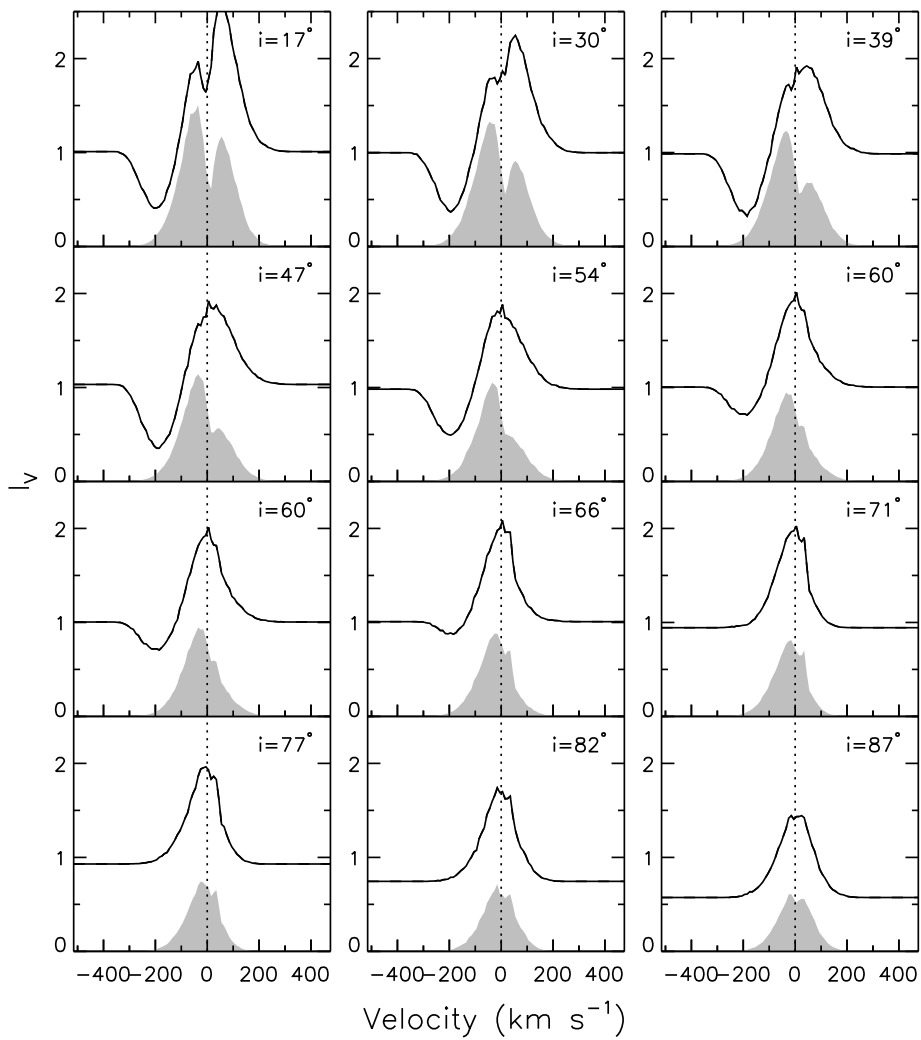}
\figcaption{ POLAR STELLAR WIND: Profiles for the same polar wind
with disk shadowing as Figure~\ref{f.stellarw_pole}, emerging within $60^o$ of the pole
with $r_i = \rho_T= 3.0 R_*$ except that
all profiles are for low turbulence ($T_L$), all include in-situ emission
($E$ = shaded area), and a much finer grid of viewing angles is shown,
stepping from $i =17^o- 87^o$.
\label{f.stellarw_pole_all}}
\end{figure}

All of the previous stellar wind models assume the wind is
spherically symmetric. {\it However, a more likely scenario for an
accretion-powered stellar wind is that it will emerge primarily
from polar latitudes on the star, above regions where stellar
magnetic field lines couple to the disk}. To assess the effects
of a restricted range of latitudes for the emergence of the
wind in the presence of disk shadowing, we show in
Figures~\ref{f.stellarw_pole} and \ref{f.stellarw_pole_all} profiles for the
case of a wind emerging only within $60^o$ of the stellar poles.
In this simulation the wind origination radius is expanded to $r_i =
3R_*$ and the disk truncation $\rho_T$ is set equal to the wind
origination radius $r_i$. The
strength of the in-situ emission in this case is double that in
previous ones to compensate for the smaller volume in the polar wind.

As expected, in the pure scattering case for the polar wind model
shown in Figure~\ref{f.stellarw_pole},
the subcontinuum blue absorption in the emergent profile remains
deep and broad at $i<60^o$, becomes half as strong at $i=60^o$,
rapidly diminishes as $i$ increases further, and disappears at
$i>71^o$. When the blue absorption is prominent the overall
profile morphology is still P-Cygni like in the low turbulence
case. However in the high turbulence case the profile is
significantly modified at small viewing angles, where the red emission
above the continuum is much less conspicuous. This results not from
disk shadowing (pure scattering profiles calculated with
$\rho_T=1.5~r_i=4.5~R_*$ are nearly the same) but from the
absence of wind gas at $90^o\ge \theta >60^o$. While at low
viewing angles the absorption of the stellar continuum (e.g. at $i=17^o$)
remains the same as in the case of $\theta_w=90^o$, the
absence of wind gas at $90^o\ge \theta >60^o$ contributes no
scattered photons from that sector. This contribution would, at
$i=17^o$, be centered about $v=100$ \kms\ (see
Figure~\ref{f.stellarw_trdisk} for $\theta_w=90^o$, $r_i=1.5$,
and $\rho_T=1.5~r_i$).  This effect is less dramatic in the
low turbulence case where the emission part of the emergent
profile, though weaker in comparison with the case of
$\theta_w=90^o$, is still conspicuous. This is because the
contribution to scattered photons into $i=17^o$
from absorption by the $90^o \ge \theta >60^o$ sector now
centers about $v=0$ \kms. Also, when the turbulence is weak
the velocity gradient is the dominant factor in radiative
transfer, and the larger velocity gradient along $\hat{r}$ means
that the probability of a photon escaping along $\hat{r}$ or
$-\hat{r}$ is higher than that along a direction $\perp$ to
$\hat{r}$. So a photon absorbed in the $90^o\ge \theta >60^o$
sector is less likely to be scattered into $i=17^o$. These
characteristics can also be deduced from the in-situ emission
profiles. Comparing the shapes (not magnitudes, because of the
different normalization factors) of the shaded profiles at
$i=17^o$ between Figures~\ref{f.stellarw_pole} and
~\ref{f.stellarw_trdisk} reveal the contribution from the
$90^o\ge \theta> 60^o$ sector to the in-situ emission profiles.

Additional profiles for the same polar wind model with
low turbulence and in-situ emission
are shown in Figure~\ref{f.stellarw_pole_all} but
with a finer grid of viewing
angles to illustrate more clearly how the
profile morphology changes with inclination. Looking at the in-situ
emission profile by itself first (shaded areas), it has a blue centroid except
at $i=87^o$. With $\rho_T=r_i$, the degree of blue-shift is
highest at $i\sim 60^o$ and decreases as $i$ moves toward $0^o$
or $90^o$. Clearly, as $\rho_T$ decreases toward $1.0$ the
degree of blue-shift increases more rapidly with more pole-on
views. For the same $\rho_T/r_i$, the in-situ emission profile
for $\theta_w=60^o$ is more blue-shifted than the corresponding
one for $\theta_w=90^o$. When the in-situ emission and the
scattering of the stellar continuum are included together the
combined profile shows that the part above the continuum changes
from having a red centroid to a blue centroid. This figure
clarifies how a polar wind generates a wide range of emergent
profiles, including some with no blue-shifted
absorption, and illustrates how the presence of disk shadowing
generates emission with blue centroids for most viewing angles.

\begin{figure}
\plotone{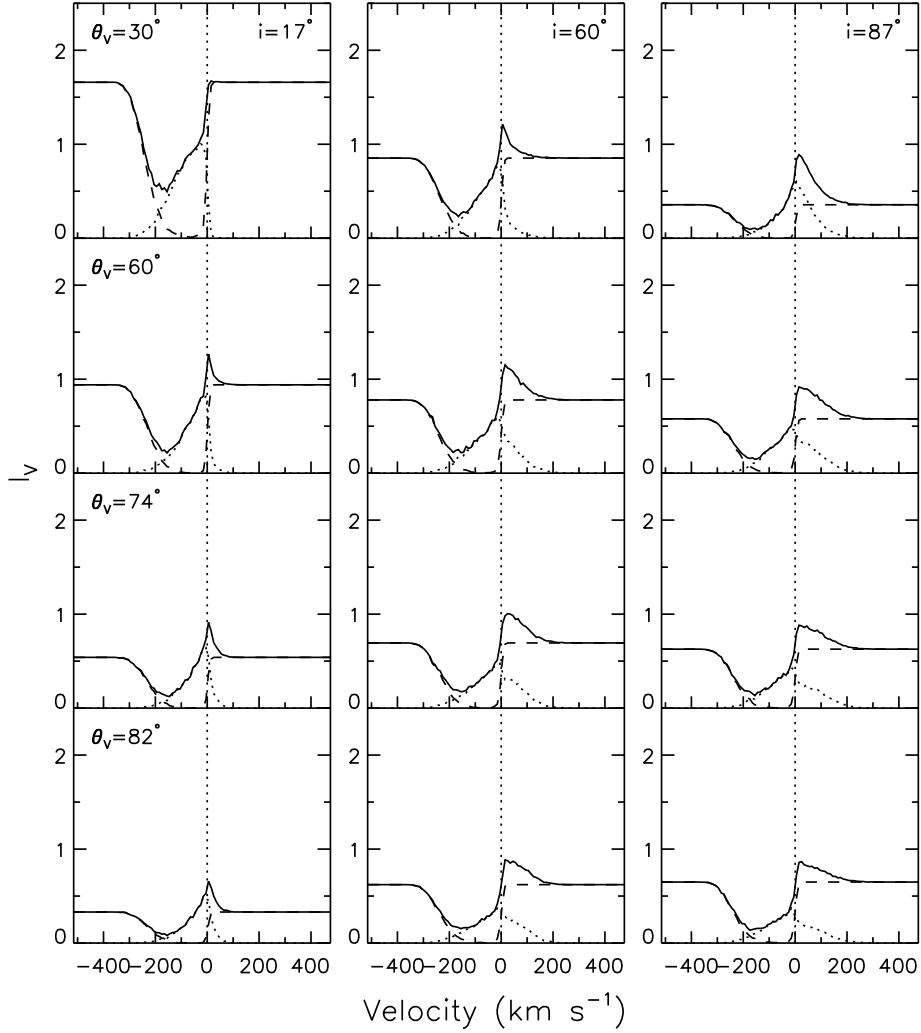}
\figcaption{ STELLAR WIND: Pure scattering profiles of a stellar
wind with the incident photons arising from a continuum source
restricted to an axisymmetric ring at polar angle $\theta_v$ on
the star with width $\Delta cos\theta=0.1$. All cases are for low
turbulence, $T_L$, and $\rho_T= 1 R_*$. The 4 rows represent 4
rings with $\theta_v=30^o$, $60^o$, $74^o$, and $82^o$. Vertical
columns denote viewing angles $i=17^o$, $60^o$, and $87^o$,
measured from the disk normal. The vertical scale illustrates
how the observed veiling continuum level depends on $\theta_v$ and $i$.
Its value relative to the stellar continuum (set to
1 at $i=0^o$) is obtained by multiplying the plotted value by the factor $0.1\times
(e^{1.33/T_*}-1)/(e^{1.33/T_v}-1)$, where $T_*$ and $T_v$ are
the stellar and veiling continuum temperature, in units of
$10^4~K$, respectively.
\label{f.stellarw_veil}}
\end{figure}

We also explored one other situation for the stellar wind models,
motivated by the fact that several CTTS with stellar wind
signatures in the \heir\ profiles have $1 \mu m$
veilings exceeding 1.0. To assess if scattering of photons from
a continuum source that is localized to rings on the star is
significantly different from scattering a continuum originating
from the entire stellar surface, we have examined the case of
photons originating solely from an isotropically radiating
continuum located as an axisymmetric ring at polar angle
$\theta_v$ of the star and of angular width about $\theta_v$
given by $\Delta cos\theta=0.1$. This simulation was carried out for a
disk reaching to the stellar surface, $\rho_T=1$, and the low
turbulence case. Figure~\ref{f.stellarw_veil} shows the
resulting line profiles, ordered from top to bottom, for
$\theta_v=30^o$, $60^o$, $74^o$, and $82^o$, respectively. The
$\theta_v=60^o$ profiles are almost identical to the ones formed
from scattering of a continuum emerging from over the whole
star. The $\theta_v =30^o$ profiles exhibit the most noticeable
difference, primarily because the observed level of this veiling
continuum is much more strongly dependent on viewing angle. When
this level is normalized to the observed stellar continuum
level, the noticeable difference lies mostly in the part of the
emergent spectrum above the continuum. This difference is not
large and, with neither $\theta_v$ nor $i$ being known in
practice, it is reasonable to take scattering of the stellar
continuum alone as being representative even when the veiling
continuum is substantial.

In sum, the profile characteristics for stellar winds in the presence of disk
shadowing are considerably more varied than those from disk winds.
Both stellar and disk wind profiles are compared to observed
ones in the next section.

\section{COMPARISON OF MODEL AND OBSERVED PROFILES}

Our profile simulations for a range of disk and stellar wind
configurations were motivated by the recent survey of \heir\ in
38 CTTS (EFHK). They take advantage of the significant
contribution from scattering in this feature, unlike previous
spectral diagnostics of inner winds in accreting systems. The
comparisons necessarily focus on profile morphology, not
intensity, which would require knowledge of the excitation
conditions in the helium wind. This approach is helpful in
determining whether the dominant signature of the inner winds in
accreting T Tauri stars points to winds arising from the star or
from the disk.

We highlight only general characteristics when comparing the data
to our schematic model profiles in the absence of other relevant
factors, particularly the relative contribution between
scattering of continuum photons and in-situ emission. Among the
ensemble of disk wind models, there is a strong dependence of
the line profile on viewing angle, ranging from a pure emission
profile at $i<\theta_w$ to a line with a prominent absorption
for $i>\theta_w$. The absorption feature is blueshifted with a
fairly sharp redward edge, always narrow in comparison with the
full range of wind velocities, and with a centroid becoming less
blueshifted as $i$ increases. For pure scattering, the emission
above the continuum is very weak, and has a broad flattened
shape from the combined rotational and poloidal motions. The
aspect dependence of the emission morphology is more clearly
delineated when in-situ emission is included, transitioning from
mostly blueshifted in pole-on views to a more centered profile
with a central depression and double horned structure in more
edge-on views.

The stellar wind models show a greater variety of profile
morphologies, arising from the effects of disk occultation,
turbulence, and wind opening angle along with the possibility of
an in-situ emission contribution to the pure scattering profile.
Key predictions from the ensemble of stellar wind models are i)
a broad range, from highly negative to highly positive values,
of the {\it net} equivalent width; ii) blue absorptions at all
inclinations for spherical winds but only for lines of sight
inside or close to the wind opening angle for polar winds; iii) for the part
of the profile below the continuum, absorption can occur over a
broad range of possible depths and widths; iv) for the part of the profile
above the continuum, emission is single peaked with centroids
that can range from blueward to redward of the stellar velocity.
Another characteristic of a stellar wind is that the blue absorption
always reaches to the blueward edge of
the profile, as it is very difficult to produce emission
blueward of the absorption, even in the presence of strong
in-situ emission. This is because the in-situ emission profile
declines with increasing blueward velocity faster than the
corresponding shallowing of the absorption profile, which is
likely to be trough-like once the line opacity exceeds unity.

\subsection{Assessment of Disk and Stellar Wind Candidates from Profiles
with Blue Absorption}

\begin{figure}
\epsscale{1.0}
\plotone{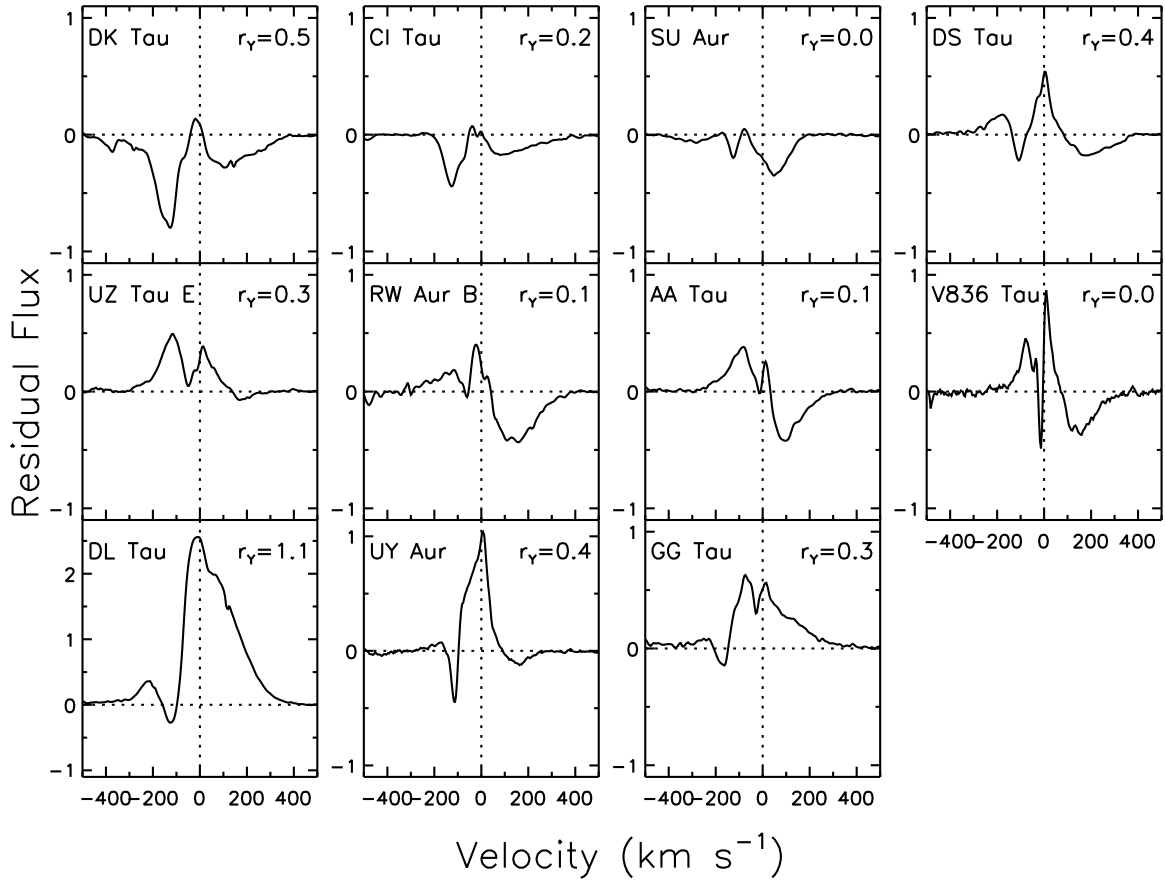}
\figcaption{ Observed profiles of \heir\ for 11 CTTS
(from EFHK) with blueshifted absorption resembling the
disk wind models. Sorting is by profile morphology.
\label{f.ctts1}}
\end{figure}

\begin{figure}
\plotone{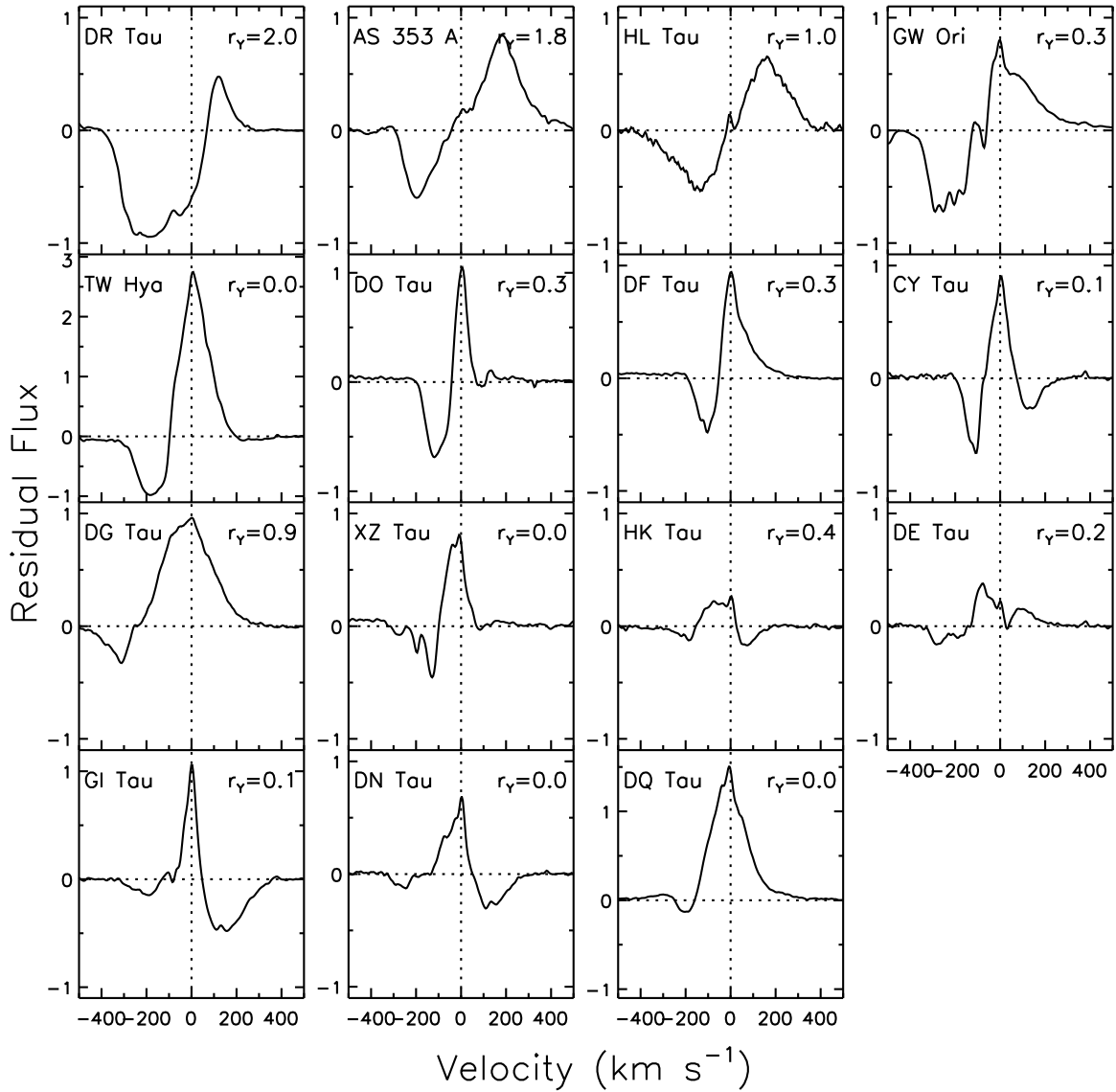}
\figcaption{Observed profiles of \heir\ for 15 CTTS
(from EFHK) with both blueshifted absorption and emission features
resembling the stellar wind models. Sorting is by profile morphology.
\label{f.ctts2}}
\end{figure}

Our initial comparison of model and observed profiles focuses on
those \heir\ profiles with blue absorption features, a clear
signature of an inner wind. Figures~\ref{f.ctts1} and
~\ref{f.ctts2} show the 26/38 \heir\ profiles from EFHK with
strong enough blue absorptions to compare to our wind models,
where the former are identified as having blue absorptions
resembling disk winds and the latter stellar winds. Some of the
observed profiles have a close resemblance to model profiles
(e.g. DK Tau and CI Tau for disk winds, HL Tau and DF Tau for
stellar winds), making the categorization straightforward.
Others are less clear (eg. DL Tau and DQ Tau) and for these we
use both (1) the presence of emission blueward of the blue
absorption and (2) the shape of the transition between the blue
absorption and the emission to its redward edge to guide our
selection in ambiguous cases, as described below.

Figure~\ref{f.ctts1} assembles helium profiles from 11 CTTS from
the EFHK survey with blueshifted absorption that we identify as
disk wind candidates: DK Tau, CI Tau, SU Aur, DS Tau, UZ Tau E,
RW Aur B, AA Tau, V836 Tau, DL Tau, UY Aur, GG Tau (sorted by profile
morphology). The majority (9) also show redshifted absorption that presumably
arises in a magnetospheric funnel flow, but these appear to be
separately formed, as it does not appear that either the blue or
red absorption is filled in by the scattered photons from the
other flow. Directing attention to the blue sides of the line
profiles, there are three stars with profiles closely resembling
those of the disk wind models with no in-situ emission and
$i>\theta_w$, {\it i.e.} narrow subcontinuum blue absorption and
little or no emission. These are DK Tau, CI Tau and SU Aur,
shown in the top row of Figure~\ref{f.ctts1}. The other 8 stars
with blueshifted absorption resembling predictions for a disk
wind also have helium emission, but this emission does not have
the character expected from a disk wind, which would be broad,
flattened, and possessive of rotational peaks for viewing angles
large enough to witness blue absorption ($i > \theta_w$). These
11 profiles thus suggest that the disk wind is not an important
source of helium {\it emission}, which must arise from another
source. In a couple of objects (V836 Tau, DL Tau) the depth of
the absorption indicates that not only the continuum but also a
part of the line emission is absorbed, suggesting the source of
emission is interior to the absorbing disk wind. In Section 3.2 we
argue that the helium emission likely arises via in-situ emission in a
stellar wind, and the narrow blue absorptions we ascribe to the disk
wind cut into this stellar wind emission.
For several stars the blue absorptions cutting
into the emission are narrower than those in
our disk wind model profiles (e.g., V836 Tau, UY Aur) but this
could result by simply narrowing the range between $\rho_i$ and
$\rho_f$ for the helium disk wind.

What is shared by all the profiles in the disk wind category is
that when helium emission is present, it extends blueward of the
narrow, shallow blue absorption feature. In stellar wind models,
which can also produce narrow and shallow blue absorption when
in-situ emission is strong, the blue absorption always reaches
to the blueward edge of the profile. In contrast, for a disk
wind the narrow absorption may be located anywhere blueward of
line center, however, it is not likely to be strong near the
wind terminal velocity. This is because although the wind
poloidal velocity has a large line-of-sight component for
viewing angles near the wind opening angle, $\theta_w$, the
sight lines toward the stellar surface will intersect the disk
wind at large distances from the disk plane where the gas
density and excitation conditions are less likely to favor a
high opacity in the \heir\ transition. Thus if there is obvious
emission extending blueward of the blue absorption we
categorize the absorption as arising in a disk
wind, with the emission arising from a second source.

A second criterion guiding our selection of blue absorption from
disk vs. stellar winds is the shape of the
transition between the blue absorption and the
emission to its red-ward edge, which is different in the two
models. In a stellar wind with pure scattering the blue
absorption gives way to the red emission smoothly, like a
slanted $\it S$ written from left to right. Even with the
addition of in-situ emission the emergent profile shows a smooth
transition between the blue absorption and the emission
component. In contrast, for the disk wind the pure scattering
profile has a distinct absorption feature with little emission
on either side, and when an in-situ emission profile is added
there is a two-component appearance, as though the absorption
feature had a ``bite'' taken out of a separate emission feature.

With these additional criteria in mind, 15 stars from the EFHK \heir\ survey, assembled in
Figure~\ref{f.ctts2},  are identified as stellar wind candidates:
DR Tau, AS 353A, HL Tau, GW Ori, TW Hya, DO Tau,
DF Tau, CY Tau, DG Tau, XZ Tau, HK Tau, DE Tau, GI Tau, DN Tau, and DQ Tau
(sorted by profile morphology). Only 4 of these also show
subcontinuum red absorption from a funnel flow, which again we assume
does not impact the blue absorption from the wind.
The profiles among the stellar wind candidates are quite varied.
Their net equivalent widths range from highly negative (DR Tau)
to highly positive (DG Tau) and their emission centroids range
from redshifted ( AS 353A, GW Ori, DF Tau) to blueshifted (HK
Tau, XZ Tau, DN Tau). In all cases the blue absorption extends to
the blue edge of the profile, although it ranges from very broad and deep
(DR Tau, AS 353A, HL Tau, GW Ori) to rather narrow and shallow
(DG Tau, HK Tau, GI Tau, DN Tau).

In contrast to the disk wind candidates, where disk winds account
only for the absorption but not the emission component of the
line, here the array of combined absorption and emission
components are well reproduced by the stellar wind models, when
the effects of disk shadowing, in-situ emission and turbulence
are considered. Some profiles resemble those predicted from pure
scattering, although they also require disk shadowing, where
$\rho_T \le r_i$, in order to create large negative equivalent
widths (DR Tau), and/or turbulence in order to reproduce red
emission peaks (DR Tau, AS 353A, and HL Tau). The contribution
of in-situ emission is required for stellar wind candidates with
a net positive equivalent width (e.g. DG Tau, TW Hya) and a
combination of in-situ emission and disk shadowing is necessary
to produce blue emission centroids (e.g. DG Tau, XZ Tau, HK Tau,
DN Tau).

We recognize that some of our categorizations are not clear-cut
cases for disk or stellar winds, especially objects like DQ Tau,
where the presence of emission blueward of the blue absorption
may or may not be present, or GG Tau, where absorptions from
both stellar and disk winds may be present. With this caveat,
our comparison of blue-shifted absorptions in observed \heir\
profiles from EFHK with those predicted from disk and stellar
wind models suggest a roughly comparable number of stars with
disk wind and with stellar wind signatures. We anticipate that a
more definitive assessment will be possible when simultaneous \heir\
and \heopt\ profiles are available, as discussed in the
following section . A summary of the profile assignments are:

\begin{list}{}
\item{$\bullet$} 11/38 CTTS (29\%) with blue absorption at \heir\
 resemble those expected for disk winds. Of these
8 also have helium emission, but with a morphology
inconsistent with formation in a disk wind viewed at an angle
exceeding the opening angle of the wind, as required for
formation of their blue absorption (Figure~\ref{f.ctts1}).
\item{$\bullet$} 15/38 CTTS (39\%) with blue absorption at \heir\
have profiles consistent with formation in stellar winds
(Figure~\ref{f.ctts2}). Here both
absorption and emission components of the profiles can be accounted
for by the stellar wind models when disk
shadowing and in-situ emission are included.
\end{list}

At least one other region besides disk and stellar winds is
clearly contributing to the formation of \heir\ -- the
(presumably) magnetospheric accretion columns channeling
infalling material from the inner disk to the star, signified by
redshifted absorption. Redshifted
subcontinuum absorption is seen in 47\% (18/38) of the CTTS in the EFHK
survey, distributed such that 9 are seen among the 11 stars
with disk wind profiles, but only 4 among the 15 stars with
stellar wind profiles. The additional 12/38 (31\%) CTTS from EFHK
that lack blue absorptions are shown in
Figure~\ref{f.ctts3}. Their characteristics are identified as:

\begin{list}{}
\item{$\bullet$}
4 stars (CW Tau, HN Tau, RW Aur A, BP Tau) where helium is entirely in
emission, with a blue centroid but no subcontinuum absorption.
 \item{$\bullet$}
4 stars (FP Tau, GK Tau, UX Tau, YY Ori) with little helium
emission but subcontinuum absorption that is at rest relative to
the star, in two cases penetrating up to 50\% of the continuum.
Two also have redshifted subcontinuum absorption.
\item{$\bullet$}
2 stars (BM And, LkCa 8) with simple reverse P Cygni profiles (and thus
redshifted subcontinuum absorption).
\item{$\bullet$}
2 stars (DD Tau, UZ TauW) with helium too weak to render an analysis.
\end{list}

While neither disk winds, stellar winds, nor funnel flows are good
candidates for the mysterious deep central absorption feature
seen in 4 stars in Figure~\ref{f.ctts3}, we consider it likely that
the remaining features in the helium lines in the EFHK survey
can be accounted for from a combination of one or more of these
three phenomena. In the next section we further explore the \heir\
profiles for clues to the presence of disk winds, stellar winds
or magnetospheres.

\subsection{In-Situ Emission and Blue Centroids  }

In the previous section we compared observed \heir\ profiles to
those predicted from disk or stellar wind models and concluded
that disk wind signatures, when present, were signified only by
the properties of the blue absorption, but not by the emission
component of the helium profile. In contrast, for profiles
attributed to stellar winds both their blueshifted absorption
and their emission above the continuum could be accounted for
with varying degrees of disk shadowing, in-situ emission and turbulence. We
examine in this section the possibility that the helium emission
in most CTTS arises in a stellar wind, whether or not the
profile also shows blue absorption from this wind. The
``emission component'' of $\lambda 10830$ will be a combination
of scattering plus true emission arising in-situ.
The relative proportion between scattering and
emission cannot be identified from profile morphology alone
except in extreme cases. If the net equivalent width is zero or
negative, it implies scattering is dominant with some
occultation from disk/star or if it is very positive, a
significant amount of in-situ emission must be present along
with the scattering.

\begin{figure}
\plotone{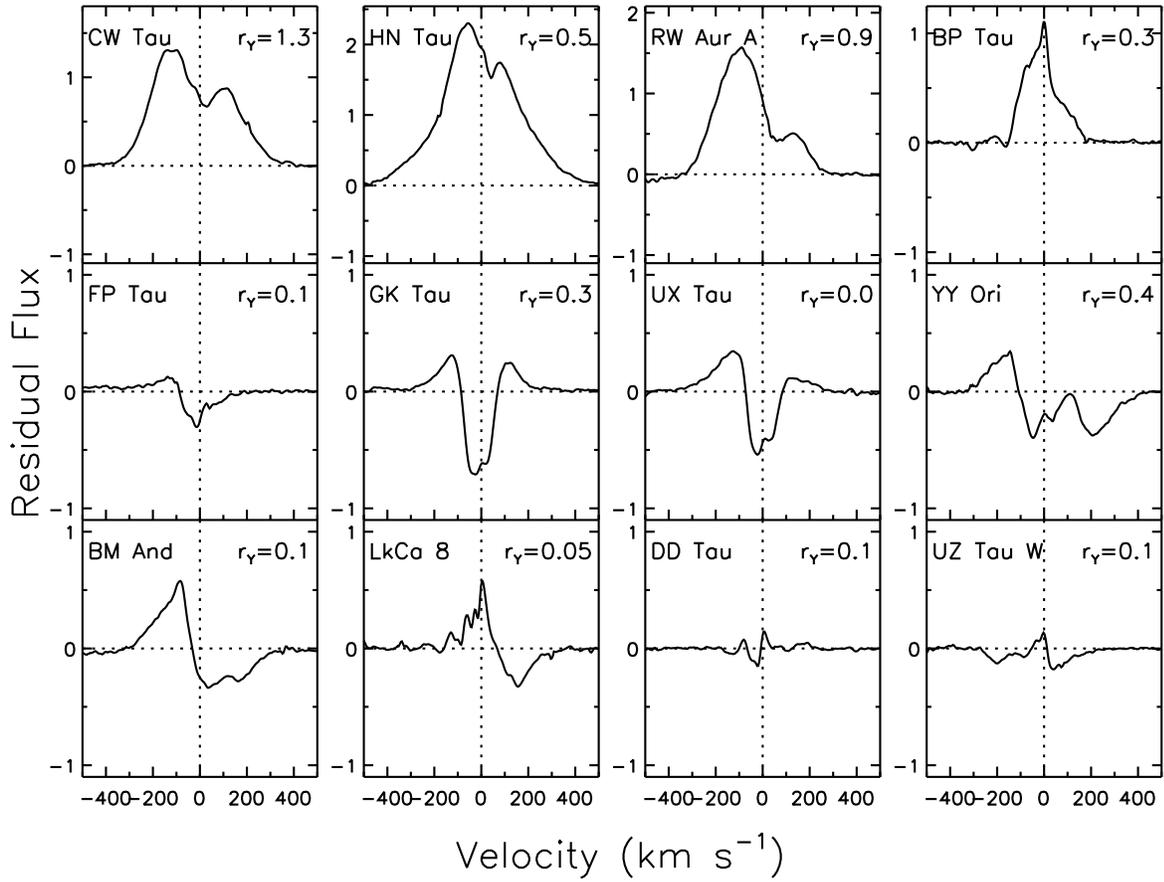}
\figcaption{Observed profiles of \heir\ for 12 CTTS
(from EFHK) lacking defining subcontinuum blueshifted absorption to
permit classification as disk or stellar winds. Sorting is by
profile morphology.
\label{f.ctts3}}
\end{figure}

Of particular interest are the 4 stars in Figure~\ref{f.ctts3}
showing \heir\ solely in emission, with no blue- or red-shifted
subcontinuum absorption. For these, the helium emission is
single-peaked, has a blueshifted centroid, and, in 3 of them,
also extends blueward in excess of 300 \kms. Single-peaked
emission with blueshifted centroids can arise in stellar winds
with occultation from the disk/star (see Figures
\ref{f.stellarw_disk},~\ref{f.stellarw_pole}) or in
magnetospheric funnel flows. In the case of the funnel flows,
such profiles will result if the line in question has similar
excitation conditions along the full arc of the magnetic
accretion column so the full range of infall velocities can
contribute to the profile \citep{mch01,bek}.

In the case of stellar winds, one possibility for producing a pure
emission profile is for the in-situ emission to be so strong
that it overwhelms the scattering contribution so no
subcontinuum absorption will be seen. This does not appear to be
the case for \heir\, however, since the observed peak flux
density exceeds the continuum flux density by an amount that is
only comparable to the latter. In the spherical stellar
wind model profiles with in-situ emission (2nd rows of Figures
\ref{f.stellarw_nodisk},~\ref{f.stellarw_disk}), even if the
in-situ emission contribution is doubled beyond what is shown,
the emergent profile will still show a blue absorption, albeit
shallower and more displaced to the blue, while the peak flux
density will be enhanced up to 3.5 times the continuum value.
This difficulty in producing a pure emission profile via strong
in-situ emission arises because the in-situ emission profile
declines with increasing blueward velocity faster than the
corresponding shallowing of the absorption profile, which is
likely to be trough-like once the line opacity exceeds unity.

A more likely means of generating \heir\ entirely in emission in
a stellar wind is one where the wind emerges from latitudes
towards the pole and is absent at latitudes close to the
equator. From our model profiles for a polar stellar wind in
Figure~\ref{f.stellarw_pole_all} it can be seen that as the
viewing angle increases from $i=0^o$ to $90^o$ the combined
profile of scattering plus in-situ emission progresses from P
Cygni-like to a profile with stronger emission and a more
displaced, weaker blue absorption, and then to a purely emission
profile that is blueshifted in the presence of disk occultation,
except for the most extreme edge-on views. Another scenario
might be an absence of spherical symmetry in the gas properties
and/or excitation conditions. Since the absorption of the
stellar continuum arises from the column of gas directly in
front of the stellar surface while both the scattered photons
into the line of sight and the in-situ emission come from a much
larger region, inhomogeneities/variations in the physical
conditions will likely affect the absorption part more and may
produce a weaker absorption, more critically at the higher
velocities. Time monitoring of the \heir\ profile can test this
possibility.

Given that single peaked helium emission with a blueshifted centroid
could signify either stellar winds or magnetospheric funnel flows
(with the caveats outlined above for each), we argue
in favor of a stellar wind for the origin of helium emission over
that from a magnetospheric funnel flow in the 3 following points.

1. Among the 15 stars in Figure~\ref{f.ctts2} with blueshifted
absorption resembling stellar winds, there is a continual
progression of \heir\ profiles from absorption dominant to
emission dominant. As shown in the previous section, a radially
accelerating stellar wind readily produces the characteristics
of the deep and broad blue absorption shown by these stars. In
situ emission arising in the same wind, with varying degrees of
contribution relative to the scattering profile (dependent on
physical conditions), can explain the rest of the profile
morphology. Invoking a second source for the helium emission,
i.e. a funnel flow, thus seems unnecessarily complicated when
the same wind producing the absorptions will suffice. Moreover,
a funnel flow origin for the helium emission in these 15 stars
would be difficult to reconcile with the observation that they
are considerably less likely to have simultaneous subcontinuum
red absorption from funnel flows than the rest of the sample
(27\% vs. 60\%).

2. A wind origin most readily accounts for the broad component
emission in a closely related line in the helium triplet series,
\heopt, which is the immediate precursor to \heir\ in a
recombination/cascade sequence. This line is one of the
strongest emission lines at optical wavelengths after the Balmer
and Ca II lines, and has a kinematic structure that is comprised of
distinct broad (BC) and narrow (NC) component emission (BEK).
The $\lambda 5876$ BC emission is optically thin and thus
directly probes the in-situ helium emission. Its kinematic
properties usually suggest an origin in outflowing gas, with
centroids either blueshifted in excess of 30 \kms\ or less
blueshifted but with blue wing velocities exceeding that
expected from an accretion flow (in excess of -200 \kms). If the
blue-centroid BC emission did arise in the accretion flow then
one would expect a correlation between the \heopt\ NC formed in
the accretion shock and the blue-centroid BC. In fact, BEK show
that the opposite is true. When the blue-centroid BC is strong,
the NC is absent or weak. In contrast, while the NC is strong,
the BC is absent or weak, and sometimes shows a redshift. A
simple explanation for this is that when in-situ helium emission
arises from the funnel flow, it is weak and its red-centroid
results from the fact that the excitation of $\lambda$5876
emission comes primarily from the terminus of the funnel flow in
close proximity to the hot accretion shock. Correspondingly, the
more prominent and more frequent $\lambda 5876$ blue-centroid BC
arises from a stellar wind, which will be the same flow
producing the \heir\ profile. (The absence of NC emission from
an accretion shock at \heir\ is discussed in \S 4.)

3. The \heir\ and $\lambda 5876$ transitions have upper states
that are more than 21 $eV$ above the ground state. Either a
temperature exceeding $20,000~K$ or a strong UV continuum flux
is needed for excitation. These conditions, though seemingly
challenging, must be present as the lower level of \heir\ is
excited in at least 3 different kinematic regions (namely,
stellar wind, disk wind, magnetospheric infall), through its
unmistakable absorption features. While for a stellar wind only a
small fraction of the energy required to accelerate the wind to
300 \kms\ or more need be tapped to heat/ionize the gas (perhaps
through magnetic field reconnection), for a magnetospheric funnel flow
no energy is needed for the infalling gas to free-fall to the star. Thus
there is a natural way for a wind to generate sufficient heating but no reason
to expect this from the funnel flow beyond the modest heating from
adiabatic compression of the converging flow \citep{martin}.
If the gas excitation in the magnetospheric funnel flow is caused instead by
photoionization, either from the accretion shock at the terminus
of the funnel flow or from the stellar corona or both, it is
reasonable to expect that the funnel flow near the terminus will
be more strongly excited, because of proximity and possible
attenuation of ionizing photons before reaching the beginning of
the funnel flow. The resulting spatial dependence of helium
excitation would produce a red-shifted BC from the funnel flow,
rather than a blue-shifted one (as sometimes seen by BEK in \heopt\ ).

Based on the above arguments attributing in-situ \heir\ emission
with blue centroids to stellar winds we assess that in addition
to the 15 objects with blueshifted absorptions resembling
stellar winds, at least an additional 9 objects may also have a
stellar wind contribution to $\lambda 10830$. These include the
4 stars where helium is entirely in emission (CW Tau, HN Tau,
RW Aur A, BP Tau in Figure~\ref{f.ctts3}) and another 5 stars
with a blue absorption indicative of a disk wind but
emission resembling that expected from in-situ emission in a stellar wind
 (DS Tau, UZ Tau E, DL Tau, UY Aur, GG Tau in Figure~\ref{f.ctts1}).
What is shared by these 9 stars with in-situ helium emission, in
varying degrees of intensity, is single-peaked and somewhat
blueshifted emission (assuming those with blue absorptions from
disk winds can be reconstructed by interpolating over the narrow
absorption). All but one also have blueward wing velocities
exceeding 300 \kms, well in excess of model profiles produced in
a funnel flow \citep{mch01}. Adding these 9 to the previous 15
makes a total of 24/38 (63\%) objects in the
EFHK survey as good candidates for helium emitting stellar winds.

This leaves 14 stars where we find no persuasive evidence for a
stellar wind contribution at helium. Among these, 5 have helium
profiles with net helium equivalent widths near zero, as
expected for profiles formed by pure scattering. In each of
these 5 a broad and deep redshifted absorption is offset by
comparably strong blueshifted ``emission'', so that the full
reverse P Cygni-like profile could be explained by scattering in
the funnel flow (BM And and LkCa 8 in Figure~\ref{f.ctts3} and
RW Aur B, AA Tau and V836 Tau in Figure~\ref{f.ctts1}). The
remaining 9 stars have ``emission'' components too weak to be
interpreted, either because the emission is intrinsically weak
or there is a deep central absorption feature rendering the
helium emission indecipherable.

We conclude there is good consistency in the evidence for helium
emitting stellar winds in accreting T Tauri stars between the
\heir\ and \heopt\ lines, although in the absence of
simultaneous optical and near-infrared spectra it is problematic
to draw conclusions from individual stars. It is also clear that
\heir, with its high frequency of subcontinuum blueshifted
absorption, is the more definitive diagnostic of a stellar wind.
However, while the \heir\ line is an excellent probe of the
kinematic behavior of gas in the vicinity of the star, by itself
it is not a strong diagnostic of physical conditions since to
scatter stellar photons it needs only a line opacity of greater
than or about unity. For example, if helium excitation is via
thermal collisions, then for a temperature between 20000-40000
K, the mass loss rate for a spherical stellar wind need only be
greater than or about $10^{-9} M_\odot yr^{-1}$ to reach an
optical depth of unity (Kwan et al., in preparation). Good
constraints on the physical conditions in these kinematic
regions await simultaneous observations of \heir\ and $\lambda
5876$.

\section{Discussion}

\subsection{Winds and Magnetospheric Accretion}

The high opacity of \heir\ provides extraordinary sensitivity
to both winds (from the disk and star) and funnel flows via
absorption of continuum photons. This allows for a comparison of the
prominence of each of these phenomena in relation to each other.
The \heir\ profiles from the EFHK survey show a clear tendency
for magnetospheric accretion signatures to be more
prevalent in stars with blue absorption from disk winds
(82\% of the stars in Figure~\ref{f.ctts1}) than those with blue
absorption from stellar winds (27\% of the stars in Figure~\ref{f.ctts2}).
A viable explanation for this could be an inclination effect if
both stellar and disk winds are simultaneously present along
with magnetic funnel flows and the stellar winds emerge from
polar latitudes. In that case polar viewing angles would
favor the appearance of P Cygni profiles from the stellar wind
while equatorial viewing angles would favor the appearance of
narrow blue absorption from disk winds along with red absorption
from funnel flows.

If the above scenario is correct we would expect large viewing
angles for stars with blue absorption from a disk wind.
Constraints on inclination for stars with stellar wind
signatures are less clear, since the opening angle of the wind
will also be important in determining whether blue absorption is
visible. A literature search turns up inclination estimates for
23 of the 38 CTTS in the EFHK survey, estimated from a variety
of techniques, including those based on (1) measurements of
rotation period and {\it v sini} coupled with an estimate for the
stellar radius \citep{ard02,app05}, (2) axial ratios of resolved
disks \citep{close98,simon00}, and (3) measured and/or inferred
radial velocities and proper motions of resolved microjets
\citep{lav00,lop03,hep04}. Often inclinations are not well
constrained, where different techniques applied to the same star
can yield estimates differing by $20^o -40^o$ degrees or more.
With these large uncertainties the value of matching profiles
with published inclinations is unclear. However, the published
inclinations do not indicate any serious contradictions to our
basic scenario. Among the stars with blue absorptions attributed
to disk winds the 7 published inclinations range from $40^o
-80^o$, consistent with viewing at high enough inclinations for
lines of sight through the disk wind. Dupree et al. (2005)
looked at \heir\ profiles in 2 CTTS with low inclinations ($<
20^o$), TW Hya and T Tau, and found both had P Cygni profiles
resembling stellar winds. The 4 stars with blueshifted helium
emission and no blue absorptions which we attribute to viewing a
polar stellar wind from inclinations exceeding the wind opening
angle are: RW Aur at $i=46^o$, CW Tau at $i= 49^o$, HN Tau at
$i= 60-70^o$, and BP Tau at $i= 30-40^o$ (see above references).
We hesitate to draw any firm conclusions from published
inclinations however, not only because many have large uncertainties
but also because misalignments between magnetic and rotation
axes may be present.

Another effect may also contribute to the less frequent
combination of stellar wind and magnetospheric accretion
signatures. Of the 7 stars in the EFHK survey with the highest
1\micron\ veilings ($r_Y$ on the order of unity or greater),
all have helium profiles attributed to stellar winds (blue
absorption and/or blue centroid emission), only one also has
blue absorption resembling a disk wind, and none
show redshifted subcontinuum absorption from a funnel
flow ($r_Y$ is identified for each profile in Figures~\ref{f.ctts1},
\ref{f.ctts2}, \ref{f.ctts3}).
Assuming that higher 1\micron\ veilings indicate higher
disk accretion rates, the absence of redshifted absorption and
higher frequency of stellar wind signatures with high veiling
could arise if disks with high accretion rates had their
truncation radii squeezed, resulting in smaller funnel flows
(less apparent redshifted absorption) while allowing larger opening angles
for stellar winds from polar latitudes. This is consistent with
the finding that the morphology for many of the stellar wind
profiles at \heir\ (Figure~\ref{f.ctts2}) seem to require
shadowing from disks extending within the wind origination radius. A
similar scenario was also suggested by BEK, based on the
relation between \heopt\ profile components and
veiling, where an anti-correlation between the strength of the
broad component emission from a stellar wind and the strength of
the narrow component emission from the accretion shock suggested
that in the presence of strong stellar winds accretion shock
formation is inhibited. This effect is also coupled with a
different proportionality between the strength of the \heopt\
narrow component emission from the accretion shock and the
amount of optical veiling, suggesting that optical veiling may
not be dominated by the accretion shock in the presence of
strong stellar winds. This could signify a transformation in the
magnetic field geometry in stars of high veiling, allowing the
disk to encroach well inside the corotation radius, reducing the
size of the accreting funnel flow and setting up conditions for
a wind to emerge radially from the star.

Unfortunately the \heir\ profiles do not provide the opportunity
to trace the presence of an accretion shock from magnetospheric
funnel flows through narrow component emission. This is, at
first glance, surprising. In contrast to \heopt\ profiles, many
of which show a prominent narrow component (BEK), this behavior
is not conspicuous at $\lambda 10830$ although NC emission may
be present in 6 objects (DS Tau, UY Aur, GI Tau, DN Tau, BP Tau
and LkCa 8) if the small emission spike near the stellar
velocity is a distinct and separate feature from the rest of the
profile. Thus, since the \heir\ transition is more
easily excited than the closely coupled \heopt\
transition, the paucity of narrow component emission at $\lambda
10830$ appears incongruous. We think two reasons account for
this apparent discrepancy. First, the narrow component is formed
in the post-shock region of a magnetospheric funnel flow where
the very high gas density makes it likely that both
transitions are optically thick and have nearly the same
excitation temperature. With a high excitation temperature for
both transitions the \heir\ intensity can be weaker than the \heopt\
intensity. When this is coupled to the fact that
the stellar continuum is higher at $1.083 \mu m$ than
at $5876$ \AA, the resultant emission at \heir\ from this region
will appear much less prominent. As an example, for a stellar temperature of
$3500~K$, and an excitation temperature of $15000~K$ and
$20000~K$, the \heir\ peak amplitude relative to
its local continuum will be $1/8.7$ and $1/10$, respectively, of
the \heopt\ peak amplitude relative to its local
continuum. If the veiling at $5876$ \AA, $r_V$, and that at
$10830$ \AA, $r_Y$, are significant, the above factor has to be
multiplied by $(1+r_V)/(1+r_Y)$. Second, a stellar wind, when
present, will scatter the narrow component. Occultation by the
star and the disk will diminish its observed intensity and
turbulence in the initial acceleration of the wind will broaden
its profile. This effect is much more important for the
\heir\ narrow component than for the $\lambda 5876$
narrow component, as the $\lambda 5876$ transition in the
stellar wind region has a much smaller opacity. For these two
reasons it is understandable why the \heir\ narrow
component is absent in objects with strong wind signatures, and
much less conspicuous, in comparison with the $\lambda 5876$
narrow component, in objects with weak or no wind signatures.

\subsection{Accretion-Powered Disk and Stellar Winds}

The ensemble of helium lines in CTTS suggests that both stellar
and disk winds emerge from near stellar regions at velocities of
several hundred \kms\, where the prominence of one over the other
at $\lambda 10830$ will depend in part on wind excitation conditions
and viewing angle. The simultaneous presence of a
disk and a stellar wind in an
accreting young star is not surprising. Most theoretical
treatments of disk winds assume, tacitly or explicitly, that a
wind will also emerge from polar regions on the star. In most of
these scenarios stellar winds are assumed to be ``ordinary''
\citep{shu94}, {\it i.e.} scaled-up analogs of the solar wind
where hot coronal gas generates sufficient thermal pressure to
accelerate material outward from the star. The observed x-ray
fluxes from T Tauri stars clarify that ``ordinary'' coronal
winds can at most generate mass loss rates up to $10^{-9}
M_\odot~ yr^{-1}$ \citep{dec81}. Recent observations show
that on average WTTS have x ray fluxes several times higher than
accreting CTTS \citep{feig06}, so any ``ordinary'' stellar wind
would actually be stronger in a WTTS than a CTTS. The stellar,
i.e. expanding radially away from the star, wind traced by
\heir\ is clearly not an ``ordinary'' one, since it is not
detected in WTTS and it predominates in CTTS with the highest
disk accretion rates. How such an accretion-powered radially
expanding wind is generated remains to be seen. Options for
MHD driven stellar winds summarized in \cite{jf06}
include driving by turbulent Alfven waves \citep{hart82} and
rotation \citep{hart80,love05},
although CTTS are not rotating fast enough for the latter to be
a major factor at this stage in their evolution.
Conceivably unsteady MHD phenomena that depend
on the orientation of the magnetic moments of the stellar and
disk fields, such as coronal mass ejections, e.g. \cite{matt02},
could generate winds with the right geometry.

Disk winds have recently been put forward to explain H$\alpha$
profiles in CTTS (Kurosawa et al. 2006; Alencar et al. 2005).
Although blue absorption with the characteristics of a stellar
wind is rare at $H\alpha$, blue absorption consistent with disk
winds is common. For example, CTTS $H\alpha$ profiles from the
BEK sample that closely overlaps the EFHK
\heir\ survey show only $10\%$ with blue absorption penetrating the
continuum, while $40\%$ of the objects have sharp
blue absorptions superposed on the broader line emission. In a
more diverse sample of young stellar objects \citep{reip96}
more than 50\% have such narrow blue absorptions
at $H\alpha$. As first suggested by Calvet (1997), this type of
narrow blue absorption at $H\alpha$ is likely formed in the disk
wind. However in a few CTTS, eg. DR Tau and AS 353A (see
profiles in BEK), $H\alpha$ has a distinct P Cygni character,
with broad subcontinuum absorption on the blue edge of the
emission profile, resembling profiles formed in stellar winds.
In a larger number of stars a stellar wind may be the source of
at least some of the $H\alpha$ emission, as evidenced by the prevalence of
blueward asymmetries in H$\alpha$ wing emission at velocities in
excess of 200 \kms\ (BEK) and by resolved spatial extensions in
the blue wings of $P\beta$ in some CTTS \citep{whelan}. The
rarity of $H\alpha$ or $H\beta$ profiles showing broad, blue
subcontinuum absorptions from a stellar wind is not surprising,
as the physical conditions for producing helium excitation (e.g.
temperatures on the order of 20,000 K or higher for thermal
collisions) will cause hydrogen to be mostly ionized. Thus
$H\alpha$ in the stellar wind would be more optically thin than
\heir\ but would still produce strong in-situ emission via
recombination and cascade.

Additional reasons for suspecting $H\alpha$ emission may arise
partly in the stellar wind is the difficulty Kurosawa et al.
(2006) had in reproducing both the emission and absorption
contributions to this line from a disk wind. Hybrid models that
include both magnetospheric and disk wind contributions were
used to calculate $H\alpha$ profiles with full radiative
transfer and self-consistently determine contributions from both
infall and outflow regions for a specified ratio of disk
accretion to mass outflow rates. However, in order to reproduce
the observed range of $H\alpha$ profiles, including both
emission and blue absorptions, disk wind opening angles in
violation of the Blandford and Payne requirement for
magnetocentrifugal launching had to be adopted. Furthermore the
tendency for CTTS to have $H\alpha$ equivalent widths increase
with inclination \citep{app05, wh04} could not be explained.
Instead they found an
inclination dependent equivalent width could be accounted for by
replacing the disk wind with a bipolar wind emerging from the
stellar polar regions. Within the context of the work presented
here on \heir, these results could be reconciled if the stellar
wind seen at helium is a major contributor to the $H\alpha$ {\it
emission}, while the blue absorption at $H\alpha$ arises in a
disk wind.

Warm disk wind models were also explored to account for CTTS
semi-forbidden lines in the ultraviolet, such as C III] and Si
III], which are single peaked, broad and blueshifted
\citep{castro05}. The models required a very restricted set of
assumptions to approximate the observed profiles. The profile
morphology of the semi-forbidden lines, however, is reminiscent
of the helium emission seen in both \heir\ and \heopt\, which we
attribute to stellar winds.

The relative importance of disk and stellar winds in accreting
systems is not yet clear. The new insight provided by \heir\ and
\heopt\ is that both are present in regions close to the star,
both are ``accretion-powered'', and when disk accretion rates are
high, a higher proportion of CTTS have helium profiles deriving
from stellar winds. This suggests that the geometry of the
star-disk interaction region is affected by the disk accretion
rate, which likely has consequences for the angular momentum
evolution of the star.

\section{CONCLUSIONS}

The \heir\ line in CTTS, through its propensity to absorb
continuum radiation, provides a unique diagnostic of kinematic
motions of the line formation regions close to the star. In this
paper we have concentrated on the blue absorptions and, with the
aid of theoretical models, have determined that both stellar and
disk winds are traced by this line. While blue absorptions
appear to arise in both disk and stellar winds, helium emission
appears to arise only in the stellar wind or in a few stars with
particularly deep red absorptions, possibly from scattering in
magnetospheric funnel flows. Some of the stellar wind profiles
also require significant disk shadowing to account for emission
with blue centroids, indicating that the disk must extend
interior to the radius where the wind is accelerated in these
stars.

A larger fraction of the CTTS in the EFHK sample show
stellar wind over disk wind signatures at \heir. However, in
view of the more prevalent signature of disk wind absorptions at
$H\alpha$, this probably indicates that excitation conditions
for \heir\ are more favorable in the stellar wind
than in the disk wind. Both are likely present in most accreting
stars, each contributing some, as yet unknown, fraction of the
outflowing gas in the large scale collimated outflows. The most
likely scenario is that stellar winds emerge from the
stellar polar regions above where magnetospheric funnel flows
intercept the star. Not only does this allow both to co-exist,
it seems to be required to explain some stars which appear to
have helium emission but not helium absorption from a stellar
wind.

Evidence that stellar winds traced by helium are
accretion-powered comes from their absence in non-accreting WTTS
and the fact that among stars with the highest disk accretion rates,
including some Class I sources \citep{e06}, the
stellar wind signatures are more frequent than disk wind signatures.
 This, coupled with the fact that the high accretion
rate stars also rarely show redshifted absorption at \heir\ from
magnetospheric infall and have weaker narrow component emission
at \heopt\ from magnetospheric accretion shocks, suggests the
possibility that when disk accretion rates are high, magnetic
accretion zones are smaller and favorable conditions for stellar winds are
enhanced. This may be an important aspect of the angular momentum
regulation mechanism postulated by Matt and Pudritz (2005),
where stellar winds act efficiently to carry away accreted
angular momentum when accretion rates are high.

It is not yet clear what role an accretion-powered stellar
wind plays in the formation of collimated jets or overall mass
and angular momentum loss in accreting systems.
While the \heir\ line is an excellent probe of the kinematic
behavior of gas in the vicinity of the star, and thus speaks
to the interaction of the stellar magnetosphere and the
accretion disk, by itself it is not
a strong diagnostic of physical conditions.
 Good constraints on the physical conditions in
these kinematic regions awaits simultaneous observations of
\heir\ and $\lambda 5876$. The latter line is the immediate precursor
to the former in a recombination/cascade sequence, so the
excitation conditions of the two lines are intimately coupled
and they will make an ideal pair for
evaluating the relative contribution between continuum
scattering and in-situ emission. This will constrain the
physical conditions, including line opacities, electron
densities, kinetic temperatures, and limits on the ionizing
flux, which in turn will enable identification of the wind
contribution to the profile, and provide estimates of the wind
mass loss rate in the helium emitting wind/s. The simultaneous
data will also help decompose the $\lambda 10830$ line profile
into separate contributions from scattering and in-situ
emission and may shed light on the wind opening angle and viewing angle.
With this information, the relative importance of disk and stellar
winds in accreting systems may be discernable.

{\it Acknowledgements}:
Collaboration and regular conversations
with Lynne Hillenbrand on related aspects of this
project has contributed to the outcome of this paper.
NASA grant NAG5-12996 issued through the Office of
Space Science provides support for this project. We thank an
anonymous referee for helpful comments that improved the
presentation of this manuscript.

\clearpage

\begin{deluxetable}{ccccccc}

\tablecaption{Guide to Figures for Disk Wind Models (Single Scattering) \label{t.1}}
\tablewidth{0pt}
\tablehead{ & \colhead{$\theta_w$} & \colhead{$\rho_i$} & \colhead{$\rho_f$} &
\colhead{Rotation} & \colhead{Emission} & \colhead{$i$} \\

\colhead{} & \colhead{(2)} & \colhead{(3)} & \colhead{(4)} &
\colhead{(5)} & \colhead{(6)}& \colhead{(7)} }
\startdata
Fig. 1 & 45$^\circ$ & $2 R_*$ & $4 R_*$, $6 R_*$ & $\Omega_C$, $\Omega_R$ & S   & 30$^\circ$, 60$^\circ$, 82$^\circ$ \\

Fig. 2 & 45$^\circ$ & $2 R_*$  & $4 R_*$& $\Omega_C$, $\Omega_R$ & S+E, S+2E  & 30$^\circ$, 60$^\circ$, 82$^\circ$\\
\enddata

\tablecomments{
Col. 2: wind opening angle, measured from disk normal;
Col. 3,4: range of disk radii for wind launching;
Col. 5:  angular momentum conservation $\Omega_C$, rigid rotation $\Omega_R$;
Col 6: emission component is either pure scattering (S) or pure scattering plus
in-situ emission (E), where E=5S has a magnitude comparable to the wind absorption;
Col 7: viewing angle, measured from disk normal.}
\end{deluxetable}

\begin{deluxetable}{cccccccc}
\tablecaption{Guide to Figures for Stellar Wind Models (Monte Carlo) \label{t.2}}
\tablewidth{0pt}
\tablehead{ & \colhead{$r_i$} &  \colhead{$\rho_T$} &
\colhead{Turbulence} & \colhead{$\theta_w$} &
\colhead{Continuum} &  \colhead{Emission} & \colhead{{\it i}}\\

\colhead{} & \colhead{(2)} & \colhead{(3)} & \colhead{(4)} &
\colhead{(5)} & \colhead{(6)} & \colhead{(7)} & \colhead{(8)}}

\startdata
Fig. 3 & 1.5 & $>7.5 $ & T$_L$, T$_H$ & 90$^\circ$ & star & S, S+E & 17$^\circ$, 60$^\circ$, 87$^\circ$ \\
Fig. 4 & 1.5 & 1.0     & T$_L$, T$_H$ & 90$^\circ$ & star & S, S+E & 17$^\circ$, 60$^\circ$, 87$^\circ$ \\
Fig. 5 & 1.5 & 2.25    & T$_L$, T$_H$ & 90$^\circ$ & star & S, S+E & 17$^\circ$, 60$^\circ$, 87$^\circ$ \\
Fig. 6 & 3.0 & 3.0     & T$_L$, T$_H$ & 60$^\circ$ & star & S, S+2E & 17$^\circ$, 60$^\circ$, 87$^\circ$ \\
Fig. 7 & 3.0 & 3.0     & T$_L$        & 60$^\circ$ & star & S+2E    & 17$^\circ$ through 87$^\circ$  \\
Fig. 8 & 1.5 & 1.0     & T$_L$        & 90$^\circ$ & rings & S      & 17$^\circ$, 60$^\circ$, 87$^\circ$ \\
\enddata

\tablecomments{
Col. (2): inner wind radius in units of $R_*$;
Col. (3): disk truncation radius in units of $R_*$;
Col. (4): low turbulence T$_L$ ($\Delta v_{tb}=10$~km~s$^{-1}$) or
high turbulence T$_H$ ($\Delta v_{tb}=100~(r_i/r)^2$~km~s$^{-1}$);
Col. (5): wind emergence angle from stellar pole (either spherically symmetric or polar)
Col. (6): continuum photons emerge either from the star or rings
at $30^\circ$, $60^\circ$, $74^\circ$, or $82^\circ$ from the stellar pole;
Col. (7) emission is either pure scattering (S) or includes in-situ emission (E) scaled to
be comparable to the wind absorption;
Col. (8) viewing angle, measured from disk normal.}

\end{deluxetable}

\clearpage

\clearpage

\clearpage

\clearpage

\clearpage

\clearpage

\clearpage

\clearpage

\clearpage

\clearpage

\clearpage

\end{document}